\documentclass[iop]{emulateapj}
\usepackage{hyperref}
\usepackage[flushleft]{threeparttable}

\newcommand{\tmbns}{$T_{\rm mb}$}

\newcommand{\ntdp}{$\rm N_2D^+~$}
\newcommand{\ntdpns}{$\rm N_2D^+$}
\newcommand{\dcop}{$\rm DCO^+~$}
\newcommand{\dcopns}{$\rm DCO^+$}
\newcommand{\ceio}{$\rm C^{18}O~$}
\newcommand{\ceions}{$\rm C^{18}O$}

\newcommand{\chthohns}{$\rm CH_3OH$}
\newcommand{\nthp}{$\rm N_2H^+~$}
\newcommand{\nthpns}{$\rm N_2H^+$}
\newcommand{\kms}{$\rm km\:s^{-1}~$}
\newcommand{\kmsns}{$\rm km\:s^{-1}$}
\newcommand{\jyb}{$\rm Jy\:bm^{-1}\:$}
\newcommand{\jybns}{$\rm Jy\:bm^{-1}$}

\shorttitle{A Hunt for Massive Starless Cores}
\shortauthors{Kong et al.}

\begin{document}

\title{A Hunt for Massive Starless Cores}

\author{Shuo Kong\altaffilmark{1,2}}
\affil{Dept. of Astronomy, University of Florida, Gainesville, Florida 32611, USA}
\affil{Dept. of Astronomy, Yale University, New Haven, Connecticut 06511, USA}

\author{Jonathan C. Tan\altaffilmark{1,3}}
\affil{Dept. of Astronomy, University of Florida, Gainesville, Florida 32611, USA}
\affil{Dept. of Physics, University of Florida, Gainesville, Florida 32611, USA}

\author{Paola Caselli\altaffilmark{4}}
\affil{Max-Planck-Institute for Extraterrestrial Physics (MPE), Giessenbachstr. 1, D-85748 Garching, Germany}

\author{Francesco Fontani\altaffilmark{5}}
\affil{INAF - Osservatorio Astrofisico di Arcetri, L.go E. Fermi 5, I-50125, Firenze, Italy}

\author{Mengyao Liu\altaffilmark{1}}
\affil{Dept. of Astronomy, University of Florida, Gainesville, Florida 32611, USA}

\author{Michael J. Butler\altaffilmark{6}}
\affil{Max Planck Institute for Astronomy, K\"onigstuhl 17, 69117 Heidelberg, Germany}

\begin{abstract}
We carry out an ALMA $\rm N_2D^+$(3-2) and 1.3~mm continuum survey towards
32 high mass surface density regions in seven Infrared Dark Clouds
with the aim of finding massive starless cores, which may be the
initial conditions for the formation of massive stars. Cores showing
strong $\rm N_2D^+$(3-2) emission are expected to be highly deuterated and
indicative of early, potentially pre-stellar stages of star
formation. We also present maps of these regions in ancillary line
tracers, including C$^{18}$O(2-1), DCN(3-2) and DCO$^+$(3-2). Over 100
$\rm N_2D^+$ cores are identified with our newly developed core-finding
algorithm based on connected structures in position-velocity space.
The most massive core has $\sim70\:M_\odot$ (potentially $\sim170\:M_\odot$)
and so may be representative of the initial conditions or early stages of massive
star formation.  The existence and dynamical properties of such cores
constrain massive star formation theories. We measure the line widths
and thus velocity dispersion of six of the cores with strongest
$\rm N_2D^+$(3-2) line emission, finding results that are generally
consistent with virial equilibrium of pressure confined cores.
\end{abstract}


\section{Introduction}

Massive star formation remains an important unsolved problem in
astrophysics. Here we seek to obtain improved observational
constraints on the initial conditions and early stages of the process.
If there is a universal star formation mechanism so that massive stars
($>$ 8 M$_\odot$)
are born via a scaled-up version of the low-mass Core Accretion
mechanism \citep[e.g.,][hereafter MT03]{1997ApJ...476..750M,2003ApJ...585..850M}, 
then the initial conditions, i.e., at the time just
before protostar formation, should be massive starless cores. Here the
term ``core'' is defined to be the self-gravitating structure that
will collapse to a single central rotationally supported disk that
eventually forms a single star or small $N$ multiple. Early stages of
massive star formation by this mechanism would include a low-mass
protostar undergoing relatively ordered accretion fed by
quasi-monolithic collapse near the center of a massive core. The
existence of such cores is a key difference between this model and
Competitive Accretion \citep[e.g.,][]{2001MNRAS.323..785B,2010ApJ...709...27W},
which involves fragmentation of gas into protostellar seeds with
initial masses only of order the thermal Jeans mass---typically much
less than a solar mass in the high mass surface density, high pressure
clumps where massive stars form. Note that the term ``clump'' is
defined to mean the self-gravitating cloud that eventually fragments
into a star cluster. Only later do some of these seeds accumulate
further material, fed from the collapsing clump, to become massive
stars.  Thus finding and characterizing massive starless and
early-stage cores is a key way to distinguish between massive star
formation theories.

However, since massive stars are rare, massive starless/early-stage
cores, even if they exist, would also be rare and thus typically far
away and relatively small in angular size.  Furthermore, they would
likely be surrounded by much larger quantities of cold, dense
molecular clump gas, with most mass going into lower mass stars or being
dispersed back into the diffuse interstellar medium. Finding massive
starless/early-stage cores is thus a challenging problem.

We have developed a strategy to overcome this challenge.  
We target regions based on mid-infrared 
extinction mapping of Infrared Dark Clouds (IRDCs) 
\citep[][hereafter BT09,
  BT12]{2009ApJ...696..484B,2012ApJ...754....5B}, which probes mass
surface densities up to $\Sigma~\sim~0.5\:{\rm g~cm}^{-2}$ and with
angular resolution of 2\arcsec.  This allows detailed study of the
structure of dense clumps: BT12 characterized 42 high $\Sigma$ clumps
selected from 10 IRDCs (A-J), which had themselves been chosen to be
relatively nearby and dense.  The 42 clumps were checked to make sure
they are free of 8 and $24\:\mu m$ (Spitzer-IRAC \& MIPS) sources. We
note that this method of sample selection differs from that based on
following up strong mm continuum sources and then selecting those that
are IR, including $70\:{\rm \mu m}$, dark 
\citep[e.g.][]{2012A&A...540A.113T,2015MNRAS.451.3089T,2016ApJ...822...59S}.

Our goal, which we carry out in this paper, is to search the majority
of these sources for \ntdpns(3-2) line emission. The abundance of this
species is known to increase in cold, dense conditions of low-mass
starless cores \citep[e.g.,][]{2005ApJ...619..379C,2007ARA&A..45..339B,2012A&ARv..20...56C},
where CO is largely frozen-out onto dust
grain ice mantles and thus depleted from the gas phase. The enhanced
abundance of \ntdp with respect to \nthp is relatively
well-understood from the astrochemical point of view, and we have
developed a comprehensive spin-state, gas phase reaction network to
model this deuteration process \citep[][hereafter K15]{2015ApJ...804...98K}. It
is this high abundance of \ntdp that acts as a signpost for the
presence of a starless core on the verge of collapse or an early-stage
core just after protostar formation, allowing us to find these
relatively rare locations in IRDCs.


We tested this method by observing 4 target regions centered on IRDC
clumps with ALMA in Cycle 0, detecting 6 \ntdpns(3-2) cores at
2\arcsec resolution ($\geq$ 1 from each region) \citep[][hereafter T13]{2013ApJ...779...96T}. 
The two most massive cores were found in the IRDC
clump C1: C1-N and C1-S.  We estimated the masses of the cores,
defined by projection of their 3$\sigma$ $l-b-v$ space \ntdpns(3-2)
contour, in two ways: (1) from the MIREX map, finding C1-N has
$61\pm30\:M_\odot$ and C1-S has $59\pm30\:M_\odot$ with the $\sim$50\%
systematic uncertainty due to assumed distance ($5\pm1$~kpc) and dust
opacity ($\sim$30\%) uncertainties; (2) from mm dust continuum
emission, finding C1-N has $16_7^{33}\:M_\odot$ and C1-S has
$63_{27}^{129}\:M_\odot$, with uncertainties mostly due to the adopted
dust temperature of $T = 10 \pm3$~K, together with distance and dust
emissivity uncertainties. Note that it is possible that in general the
``core'' may extend beyond the observed \ntdpns(3-2) contour, so these
may be lower limits on the core mass. On the other hand, it is also
possible the \ntdpns(3-2) structure may actually contain more than one
core, i.e., it may be resolved into two or more separate cores if
observed at higher angular resolution.

Thus of the six T13 cores, C1-S and C1-N are the most promising
examples of a massive starless/early-stage cores, i.e., with
$\ga20\:M_\odot$ that may allow formation of a $\ga10\:M_\odot$ star,
given expected outflow regulated formation efficiencies $\sim$50\%
\citep{2014ApJ...788..166Z}. C1-S appears monolithic, centrally-concentrated
in both \ntdpns(3-2) and mm continuum emission, and rounded (most
likely by self-gravity). C1-N appears to be less centrally
concentrated and potentially fragmented. Follow-up observations of
other \ntdp and \nthp lines allowed measurement of $D_{\rm
  frac}\equiv [{\rm N_2D^+}]/[{\rm N_2H^+}]$ in the cores, with values
of 0.2-0.7 \citep{2016ApJ...821...94K}. For most chemodynamical models, such
high values that are orders of magnitude greater than the cosmic
[D]/[H] ratio of $\sim 10^{-5}$, imply relatively old astrochemical
ages and thus relatively slow collapse rates, $\lesssim 1/3$ of the
rate of free-fall collapse.

Further follow-up with ALMA in Cycle 2 of the C1 region found the
presence of a very collimated protostellar outflow, traced by
$^{12}$CO(2-1), from a source within C1-S (in both position and
velocity space), so that this is most likely to be an example of an
early-stage massive core \citep{2016ApJ...821L...3T}. A second protostellar
outflow source also overlaps spatially with C1-S, although its
association with the core in velocity space is less certain. No
outflows were seen from C1-N.

T13 used the \ntdpns(3-2) line-width to study the dynamics of the
cores. For the sample of 6 sources, the velocity dispersions were on
average consistent (within a factor of $\sim 0.8$) with those expected
from virial equilibrium of the fiducial MT03 Turbulent Core
model. However, for C1-S the observed velocity dispersion is about a
factor of two smaller than the fiducial virial equilibrium
prediction. If virial equilibrium is being maintained, as would be
expected if the astrochemical age is larger than the dynamical time,
then relatively strong magnetic fields, $\sim$1~mG, are needed.

We see that larger samples of starless and early-stage cores are
needed to better test the different theoretical models. This has
motivated the observations and analysis presented in this paper.

\section{Sample Selection and ALMA Observations}\label{sec:obs}

\begin{deluxetable*}{ccccrcc}
\tabletypesize{\scriptsize}
\tablecolumns{7}
\tablewidth{0pc}
\tablecaption{IRDC Clump Targets of the ALMA Cycle 2 Observation\label{tab:obs}}
\tablehead{
\colhead{Clump$^1$} &\colhead{R.A.} & \colhead{DEC.} & \colhead{l (deg)} & \colhead{b (deg)} &\colhead{$v_{\rm LSR}$~(\kmsns)} & \colhead{$d$~(kpc)}
}
\startdata
A1 & $\rm 18^h26^m15\fs14$ & $\rm -12\arcdeg41\arcmin43\farcs4$ & 18.78675 & -0.28592 & 66 & 4.8\\
A2 & $\rm 18^h26^m19\fs04$ & $\rm -12\arcdeg41\arcmin14\farcs8$ & 18.80117 & -0.29625 & 66 & 4.8\\
A3 & $\rm 18^h26^m21\fs78$ & $\rm -12\arcdeg41\arcmin10\farcs2$ & 18.80750 & -0.30550 & 66 & 4.8\\
B1 & $\rm 18^h25^m52\fs83$ & $\rm -12\arcdeg04\arcmin52\farcs7$ & 19.28758 &  0.08083 & 26 & 2.4\\
B2 & $\rm 18^h25^m58\fs29$ & $\rm -12\arcdeg04\arcmin13\farcs5$ & 19.30758 &  0.06625 & 26 & 2.4\\
C2 & $\rm 18^h42^m50\fs45$ & $\rm -04\arcdeg03\arcmin17\farcs7$ & 28.34383 &  0.06017 & 79 & 5.0\\
C3 & $\rm 18^h42^m44\fs02$ & $\rm -04\arcdeg01\arcmin54\farcs5$ & 28.35217 &  0.09450 & 79 & 5.0\\
C4 & $\rm 18^h42^m49\fs34$ & $\rm -04\arcdeg02\arcmin27\farcs3$ & 28.35417 &  0.07067 & 79 & 5.0\\
C5 & $\rm 18^h42^m52\fs59$ & $\rm -04\arcdeg02\arcmin44\farcs3$ & 28.35617 &  0.05650 & 79 & 5.0\\
C6 & $\rm 18^h42^m54\fs37$ & $\rm -04\arcdeg02\arcmin31\farcs7$ & 28.36267 &  0.05150 & 79 & 5.0\\
C7 & $\rm 18^h42^m40\fs01$ & $\rm -04\arcdeg00\arcmin34\farcs4$ & 28.36433 &  0.11950 & 79 & 5.0\\
C8 & $\rm 18^h42^m59\fs99$ & $\rm -04\arcdeg01\arcmin33\farcs1$ & 28.38783 &  0.03817 & 79 & 5.0\\
C9 & $\rm 18^h42^m51\fs86$ & $\rm -03\arcdeg59\arcmin43\farcs3$ & 28.39950 &  0.08217 & 79 & 5.0\\
D1 & $\rm 18^h44^m17\fs05$ & $\rm -04\arcdeg02\arcmin01\farcs5$ & 28.52717 & -0.25033 & 87 & 5.7\\
D2 & $\rm 18^h44^m23\fs79$ & $\rm -04\arcdeg02\arcmin11\farcs5$ & 28.53750 & -0.27650 & 87 & 5.7\\
D3 & $\rm 18^h44^m15\fs38$ & $\rm -04\arcdeg00\arcmin50\farcs7$ & 28.54150 & -0.23517 & 87 & 5.7\\
D4 & $\rm 18^h44^m22\fs51$ & $\rm -04\arcdeg01\arcmin53\farcs5$ & 28.53950 & -0.26950 & 87 & 5.7\\
D5 & $\rm 18^h44^m16\fs43$ & $\rm -03\arcdeg59\arcmin22\farcs3$ & 28.56533 & -0.22783 & 87 & 5.7\\
D6 & $\rm 18^h44^m17\fs78$ & $\rm -04\arcdeg00\arcmin12\farcs4$ & 28.55550 & -0.23917 & 87 & 5.7\\
D7 & $\rm 18^h44^m17\fs69$ & $\rm -03\arcdeg59\arcmin26\farcs5$ & 28.56667 & -0.23300 & 87 & 5.7\\
D8 & $\rm 18^h44^m18\fs29$ & $\rm -03\arcdeg59\arcmin06\farcs2$ & 28.57283 & -0.23267 & 87 & 5.7\\
D9 & $\rm 18^h44^m18\fs74$ & $\rm -03\arcdeg58\arcmin13\farcs7$ & 28.58667 & -0.22767 & 87 & 5.7\\
E1 & $\rm 18^h43^m06\fs71$ & $\rm -03\arcdeg45\arcmin09\farcs9$ & 28.64350 &  0.13817 & 80 & 5.1\\
E2 & $\rm 18^h43^m10\fs12$ & $\rm -03\arcdeg45\arcmin15\farcs8$ & 28.64850 &  0.12483 & 80 & 5.1\\
F3 & $\rm 18^h53^m18\fs42$ & $\rm +01\arcdeg27\arcmin33\farcs9$ & 34.44383 &  0.24967 & 58 & 3.7\\
F4 & $\rm 18^h53^m18\fs51$ & $\rm +01\arcdeg28\arcmin30\farcs5$ & 34.45800 &  0.25650 & 58 & 3.7\\
H1 & $\rm 18^h57^m11\fs37$ & $\rm +02\arcdeg07\arcmin27\farcs1$ & 35.47800 & -0.31033 & 44 & 2.9\\
H2 & $\rm 18^h57^m06\fs92$ & $\rm +02\arcdeg08\arcmin20\farcs9$ & 35.48283 & -0.28700 & 44 & 2.9\\
H3 & $\rm 18^h57^m08\fs83$ & $\rm +02\arcdeg08\arcmin24\farcs3$ & 35.48733 & -0.29367 & 44 & 2.9\\
H4 & $\rm 18^h57^m06\fs88$ & $\rm +02\arcdeg08\arcmin44\farcs8$ & 35.48867 & -0.28383 & 44 & 2.9\\
H5 & $\rm 18^h57^m08\fs27$ & $\rm +02\arcdeg08\arcmin57\farcs7$ & 35.49450 & -0.28733 & 44 & 2.9\\
H6 & $\rm 18^h57^m08\fs17$ & $\rm +02\arcdeg10\arcmin51\farcs8$ & 35.52250 & -0.27250 & 44 & 2.9
\enddata
\tablenotetext{1}{Targets selected from BT12.}
\end{deluxetable*}

BT09 and BT12 studied the 10 IRDCs A-J,
selecting them from the sample of \citet{2006ApJ...641..389R} to be
particularly suitable for MIREX mapping: they are relatively nearby
(thus reducing the contribution of MIR foreground emission); show high
contrast against the Galactic MIR background; and are surrounded by
relatively simple, smooth MIR background emission.  BT12 analyzed the
detailed structural properties of 42 clumps selected from these clouds
to be dark at 8 \& 24~$\rm \mu m$, the latter evaluated from the
MIPSGAL survey \citep{2009PASP..121...76C}.  T13 observed \ntdpns(3-2) in 4
clumps, C1, F1, F2, G2, with ALMA in Cycle 0.

With ALMA in Cycle 2 (Project number: 2013.1.00806.S; PI: Tan), we
observed 32 more clumps from the BT12 sample (listed in Table
\ref{tab:obs}), focusing on IRDCs A, B, C, D, E, F and H.  
The observations were carried out during April 2015
with the 12m array in the most compact configuration.
The baselines
were from 12~m to 330~m (9-254 k$\lambda$), resulting in a synthesized
beam size of 1.5\arcsec~$\times$~1.0\arcsec and a maximum detectable
scale of $\sim$20\arcsec.
The diameter of the primary beam, which
approximately sets the field-of-view (FOV), is $\sim$26\arcsec.
The maximum scale is comparable to the FOV, and ALMA has
very good uv-coverage in the short spacings in the compact configuration.
No ACA observations were performed.
Two of the target pointings contained two BT12 clumps, so in
total there were 30 pointings in our observations (two tracks, each of
15 pointings).  Together with the 4 targets already observed, this
completes 86\% of the BT12 sample (92\% of the sources in IRDCs A-H,
which are at a range Galactic longitudes from $l = 18.8\arcdeg$ to
$35.5\arcdeg$).

Dual polarization mode was adopted for the Band 6 spectral setup.
Four basebands and seven spectral windows were used during the
observations. A Baseband 1 single spectral window was centered on
\ntdpns(3-2) (rest frequency 231.32~GHz), with a velocity resolution
of 0.05 \kmsns. A Baseband 2 single spectral window was used for a
continuum observation, centered at 231.00~GHz. The total bandwidth for
this baseband is about 2 GHz. A Baseband 3 single spectral window was
centered on \ceions(2-1) (rest frequency 219.56~GHz), with a velocity
resolution of 0.05 \kmsns.  Baseband 4 was split into four spectral
windows, including a window at 216.11 GHz for \dcopns(3-2), a window
at 216.95~GHz for CH$_3$OH($v$ $t=0$ 5(1,4)-4(2,2), with upper-state
energy of 56~K), a window at 217.10 GHz for SiO(5-4), and a window at
217.24 GHz for DCN(3-2). Each of these lines has a 0.2 \kms velocity
resolution.

This paper will focus mostly on the results from the continuum and
\ntdpns(3-2) observations, although the integrated intensity maps of
most of the other species are also presented. The SiO(5-4) data, which
probe protostellar outflows, are presented by Liu et al. (in prep).

The sample of 30 targets was divided into two tracks, each containing
15 sources. Track 1, with reference velocity of +58~\kmsns, includes
A1, A2, A3 \citep[$v_{\rm LSR}\simeq+66\:$\kmsns: these estimates are
derived from $^{13}$CO(1-0) emission from the clouds: see, e.g.,][]{2015ApJ...809..154H}, 
B1, B2 ($v_{\rm LSR} \simeq +26\:$\kmsns), C2\footnote{We note that the C2 region was studied by \citet{2015ApJ...804..141Z}, including an observation of \ntdp with a sensitivity of 0.0075 \jyb per 0.7 \kmsns. They detected several mm continuum cores in this region, most of which appear to be protostellar.},
C3, C4, C5, C6, C7, C8, C9 ($v_{\rm LSR} \simeq +79\:$\kmsns), E1, E2
($v_{\rm LSR} \simeq +80\:$\kmsns). Track 2, with reference velocity
of +66~\kmsns, includes D1, D2 (also contains D4), D3, D5 (also
contains D7), D6, D8, D9 ($v_{\rm LSR} \simeq +87\:$\kmsns), F3, F4
($v_{\rm LSR} \simeq +58\:$\kmsns), H1, H2, H3, H4, H5, H6\footnote{Also studied by \citet{2016arXiv160800009H} with ALMA Band 7 observations, detecting multiple sub-mm continuum cores, including some embedded in very narrow filamentary structures.} ($v_{\rm
  LSR} \simeq +44\:$\kmsns).
In Track 1, J1924-2914 was used as bandpass calibrator, J1832-1035 was
used as gain calibrator, and Neptune was the flux calibrator.  In Track 2,
J1751+0939 was used as bandpass calibrator, J1851+0035 was used as
gain calibrator, and Titan was the flux calibrator.
The continuum data were cleaned, while the line data were
not cleaned due to their relatively weak detections.
No self-calibration was done to the dataset.
Primary beam correction was applied before fluxes were extracted.

Our sensitivity level was set by the desire to detect massive \ntdp
cores that are similar to C1-S with $\gtrsim 5\sigma$ significance
integrating over the typical velocity range of such cores, $\sim
1$~\kmsns. We estimated this to be a sensitivity level of 30~mJy per
beam per 0.1~\kmsns, i.e., 3 times worse than that achived by T13. In
the end, Track 1 has sensitivity of 22~mJy beam$^{-1}$ per 0.1~\kmsns.
In Track 2, the sensitivity was 30~mJy beam$^{-1}$ per 0.1~\kmsns.  A
5~\kms integration results in a noise level of 21~mJy
beam$^{-1}$~\kmsns. 
The continuum sensitivity we achieve now is 0.22 mJy beam$^{-1}$,
compared with T13's value of 0.27~mJy per 2.3\arcsec $\times$
2.0\arcsec beam.

\section{Core Detection Analysis Methods}\label{sec:method}

Our main goal is to systematically identify \ntdpns(3-2) cores,
eventually presenting a rank ordering of cores via their \ntdp line
flux that extends down to relatively weak cores for which there begins
to be possible confusion with noise fluctuations. We will utilize the
information of colocation of the cores with other tracers (e.g.,
1.3~mm continuum emission and other line tracers) to help assess the
reliability of the \ntdpns(3-2) cores. For the strongest cores we will
then estimate masses from both 1.3~mm dust emission and from the MIREX
maps. Those cores with $\ga 10\:M_\odot$ and without star forming
activity are good candidates to be massive starless cores.

We start by identifying cores as connected groups of ``voxels''
(pixels in 3D position-position-velocity (PPV) space of $l-b-v$) with
flux densities of $\geq 3\sigma$ (``first threshold''), where $\sigma$
is the noise level in each voxel. 
Note that this search is done
  in the cube before primary beam correction to avoid spurious
  features at the edge of the maps; also we restrict the core finding
  to within a field of view defined by the primary beam diameter.
However, the noise level, $\sigma$, depends on the velocity resolution
to which we smooth the data. After some experimentation (discussed
below), we settle on a choice of a velocity resolution of 0.15~\kmsns
(i.e., 3 times coarser than the full resolution of the data cubes),
which has $\sigma\simeq0.020\:$\jyb for IRDCs A, B, C, E (Track 1) and
$\sigma\simeq0.030\:$\jyb for IRDCs D, F, H (Track 2). Next, searching
only within the field-of-view (FOV) of diameter of 26\arcsec (i.e.,
the half-power response diameter of the primary beam, i.e., where the
sensitivity is reduced by about 50\%), we identify all voxels above
the $3\sigma$ threshold.

The selected voxels are then denoted as ``nodes'' (as in Graph
theory). If two nodes are connected, i.e., adjacent in $l-b-v$ space,
they are connected with an ``edge.''  Once the data cube is traversed,
we have a set of nodes and edges, and thus a Graph.  Note here that
the Graph so constructed has no directional information (i.e., an
``undirected Graph''): if voxel A is connected to voxel B, so is voxel
B to voxel A. Next, using the NetworkX package
\citep{hagberg-2008-exploring}, we identify all connected components
in the graph, each of them being a core candidate.

After checking the results and comparing with the region (clump-scale)
integrated intensity maps of \ntdpns(3-2) (using a 5\kms velocity
range), we notice some potential \ntdpns(3-2) structures that are not
found by the above selection method. These structures have very few
voxels that are above the first core threshold. However, they show
continuous positive flux in velocity space. This leads us to amend our
voxel selection: for already identified $\geq3\sigma$ voxels, we
consider their neighboring voxels and count them as part of the core,
i.e., as nodes in the Graph, if they have a positive \ntdpns(3-2) line
flux. The effect of this is to add the contribution of a halo of
\ntdpns(3-2) line emission around the stronger peaks.

One can see that core selection may depend on a number of choices of
criteria, including PPV cube velocity resolution, noise selection
thresholds, and minimum angular size. We considered a variety of
velocity resolutions with which to smooth the data cubes, which have
an original full resolution of 0.05~\kmsns. Based on the observations
of T13, \ntdpns(3-2) cores show line widths (including hyperfine
broadening) as narrow as $\sim0.5\:$\kmsns. Therefore, while we have
adopted a smoothing of the data to 0.15~\kms resolution, we also
make it a requirement that the core is detected in at least two
adjacent channels. Since the {\it ALMA} data by default is Hanning
smoothed, two neighboring channels are potentially correlated. Thus, a
noise spike could potentially span across two channels, even when
smoothed to 0.15~\kmsns, causing a false detection. To eliminate such
false cores, we re-make the PPV cube with a 0.075~\kms velocity
shift in the boundaries of the velocity channels and require cores to
also be identified in this cube. The 0.075~\kms shift is more than
twice the finest channel spacing of the raw data, so these two cubes
provide a good cross check to rule out noise spike features. Finally,
we impose a condition of a minimum number of voxels that span an area
on the sky that is comparable to the synthesized beam. This condition
also helps mitigate false detections due to noise spikes.

Once \ntdpns(3-2) cores have been identified, we make additional checks
of other tracers co-located at the core position and velocity.  At the
position of an \ntdp core, we check the signal in the continuum image
and the 0th-moment images of DCO$^+$(3-2) and C$^{18}$O(2-1). Based on
the observational results from T13, all \ntdp cores have corresponding
\dcop and continuum emission. In cold starless cores, CO and its
isotopologues should suffer heavy depletion via freeze-out on to dust
grains, thus significantly reducing the flux in, e.g.,
C$^{18}$O(2-1). However, note that these checks are not used in core
selection, but only to provide additional information that can then be
used to assess the reality of the \ntdp core.

Finally, having obtained our sample of selected \ntdpns(3-2) cores we
re-examine their correspondence in the region (clump-scale) integrated
intensity maps of \ntdpns(3-2) that use a 5~\kms velocity range
(thus with a higher noise level). The strongest cores are easily
visible in this map, but many of the weaker ones do not stand out
above the noise. In addition there remain some $\geq4\sigma$ features
visible in the region maps that have not been identified as cores by
our above method (i.e., as connected structures in PPV space). Since
our focus is on the cores, we simply note the position of these
``clump-scale'' structures and carry out simple checks with other
tracers to gauge their reality. 

%
Our adopted method of finding cores, like any method, does involve
some somewhat arbitrary choices for core definition and extraction. We
have thus compared it with the ``clumpfind'' method in the yt package
\citep{2009ApJ...691..441S}. We carried out the comparison in the B1
region. When we set the minimum size as the synthesized beam size, the
search radius as the primary beam, and the threshold as 2.5$\sigma$,
then yt finds all B1 cores given by our method.  We consider that both
methods find the coherent signals in the PPV cube.  Ultimately though
the precise definition used for a core will influence the final
results. Thus when comparing, for example, to the outputs of numerical
simulations, similar analysis methods should be adopted where
possible.

\section{Results}\label{S:results}

\subsection{Overview of the 32 IRDC Clumps}\label{S:regions}


\begin{figure*}[htb!]
\epsscale{0.85}
\plotone{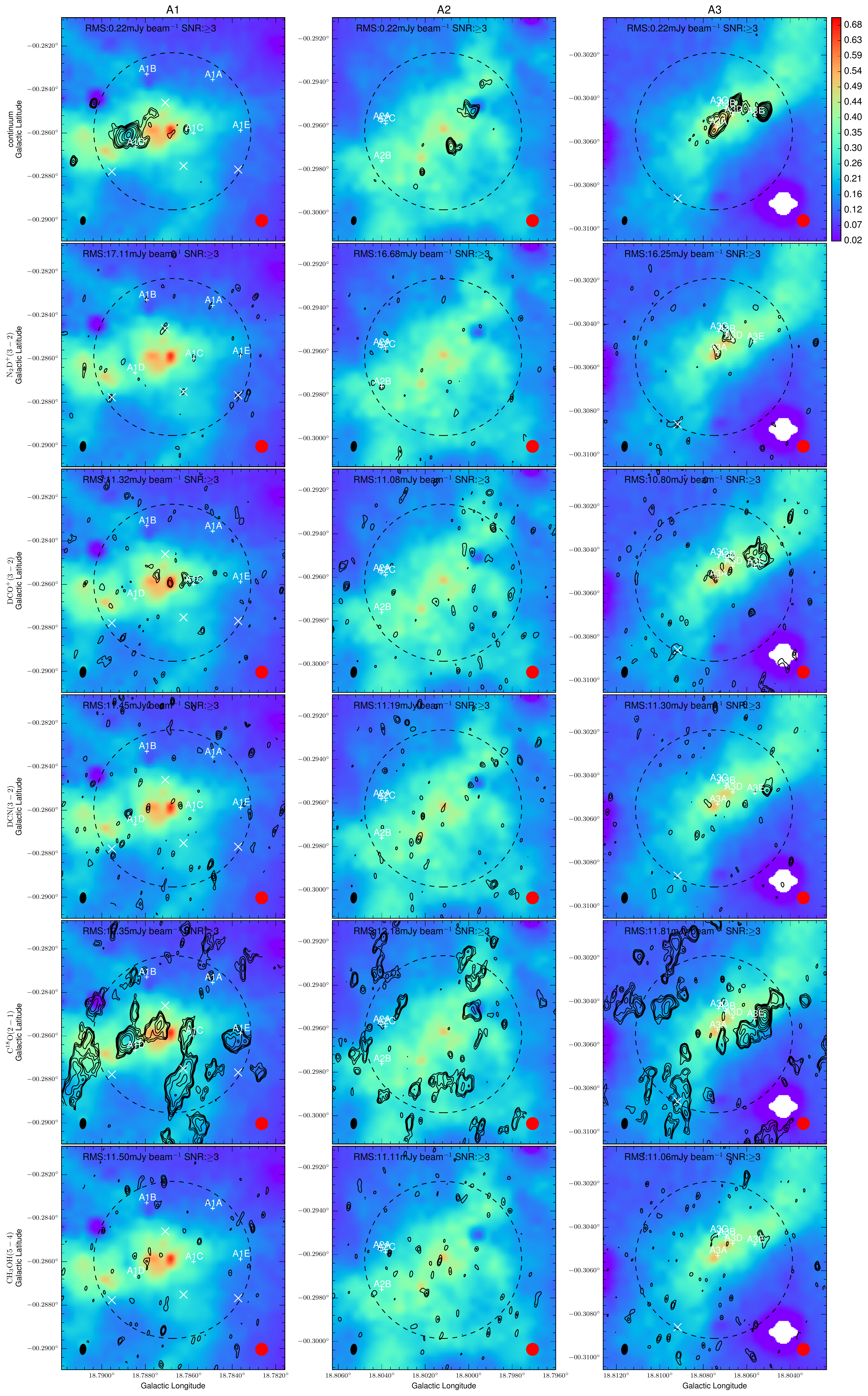}
\caption{
Summary figures for the surveyed IRDC clumps. Columns from left to
right show the results for sources A1, A2, A3 (other sources continue
in subsequent figures). The color-scale background shows MIREX mass
surface density map from BT12. The color bar at top-right corner
indicates the scale in g~cm$^{-2}$, which is preserved throughout all
30 columns.  The white regions are locations of MIR bright sources,
where the map is undefined. The ALMA survey results are shown as
contours. From top to bottom, the contours show 1.30~mm continuum,
integrated intensities of \ntdpns(3-2), \dcopns(3-2), DCN(3-2),
\ceions(2-1), \chthohns($v$ $t=0$ 5(1,4)-4(2,2)). The contour levels are
in unit of $\sigma$, starting from 3$\sigma$, 4$\sigma$, 5$\sigma$,
7$\sigma$, 10$\sigma$, 15$\sigma$, 20$\sigma$, 30$\sigma$, 40$\sigma$,
50$\sigma$, 70$\sigma$, 100$\sigma$, 130$\sigma$, 160$\sigma$,
190$\sigma$...
The $\sigma$ for 1.30 mm continuum is 2.2$\times$10$^{-4}$ \jybns. The
integrated intensity maps are made within a 5~\kms velocity range,
with a $\sigma$ of about 0.02 \jyb \kmsns. In each panel we show the
diameter of the primary beam is shown with the dashed circle and the
{\it ALMA} beam in lower left and {\it Spitzer} 8~$\rm \mu m$ beam in
lower right (that sets the resolution of the MIREX map). Each panel
also shows locations of \ntdpns(3-2) cores identified as connected
structures in PPV space with ``$+$'' signs, along with the core
names, e.g., A1A. Note that sometimes these cores do not appear as
$\geq3\sigma$ features in the integrated intensity map of \ntdpns(3-2)
in the 2nd row (see text). Sometimes $\geq4\sigma$ features are seen
in this map that are not associated with identified cores and these
are marked with ``$\times$'' signs (see text).
\label{fig:a1a2a3}}
\end{figure*}

\begin{figure*}[htb!]
\epsscale{0.95}
\plotone{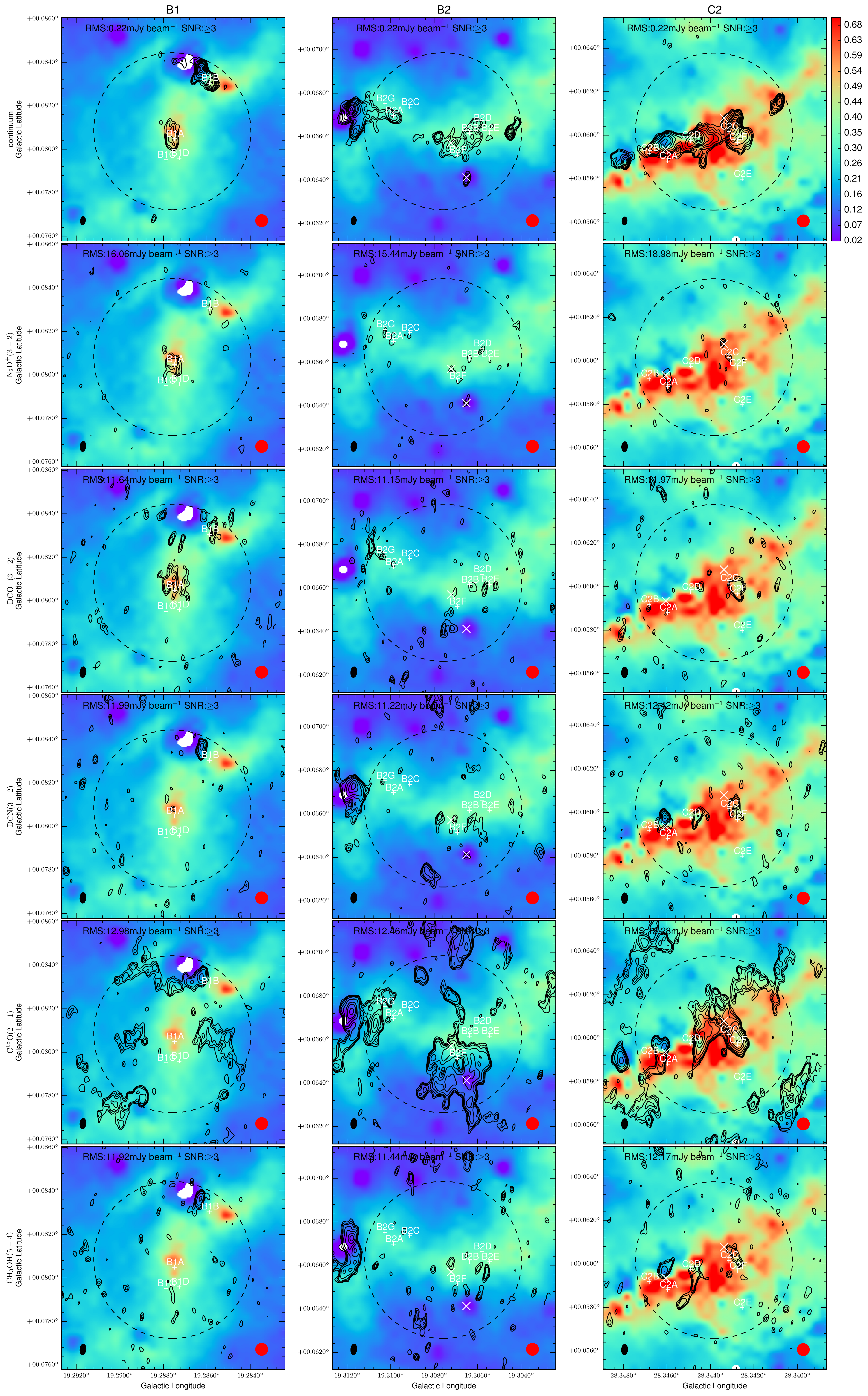}
\caption{
Same as Fig. \ref{fig:a1a2a3}, but for sources B1, B2, C2.
\label{fig:b1b2c2}}
\end{figure*}

\begin{figure*}[htb!]
\epsscale{0.95}
\plotone{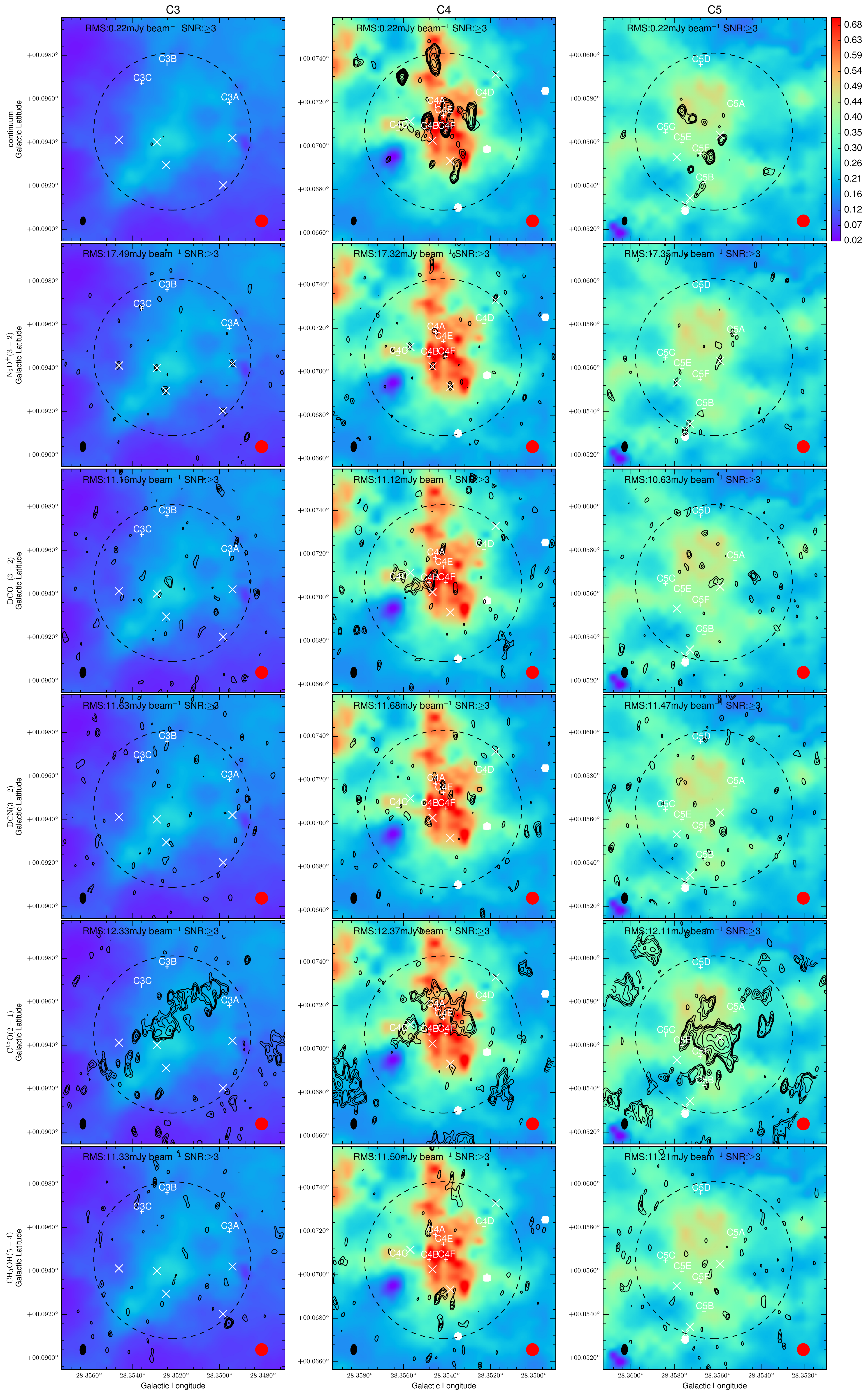}
\caption{
Same as Fig. \ref{fig:a1a2a3}, but for sources C3, C4, C5.
\label{fig:c3c4c5}}
\end{figure*}

\begin{figure*}[htb!]
\epsscale{0.95}
\plotone{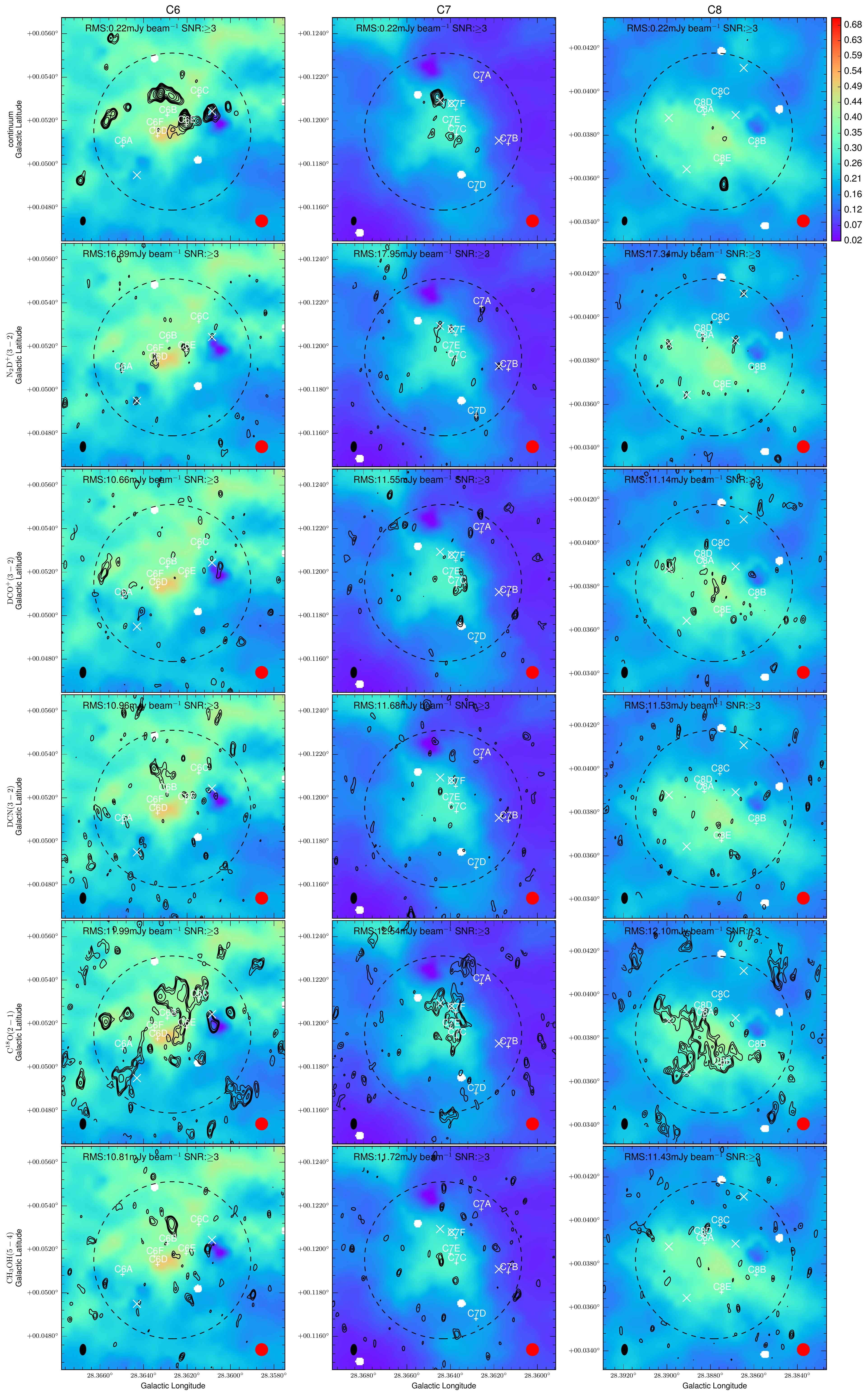}
\caption{
Same as Fig. \ref{fig:a1a2a3}, but for sources C6, C7, C8.
\label{fig:c6c7c8}}
\end{figure*}

\begin{figure*}[htb!]
\epsscale{0.95}
\plotone{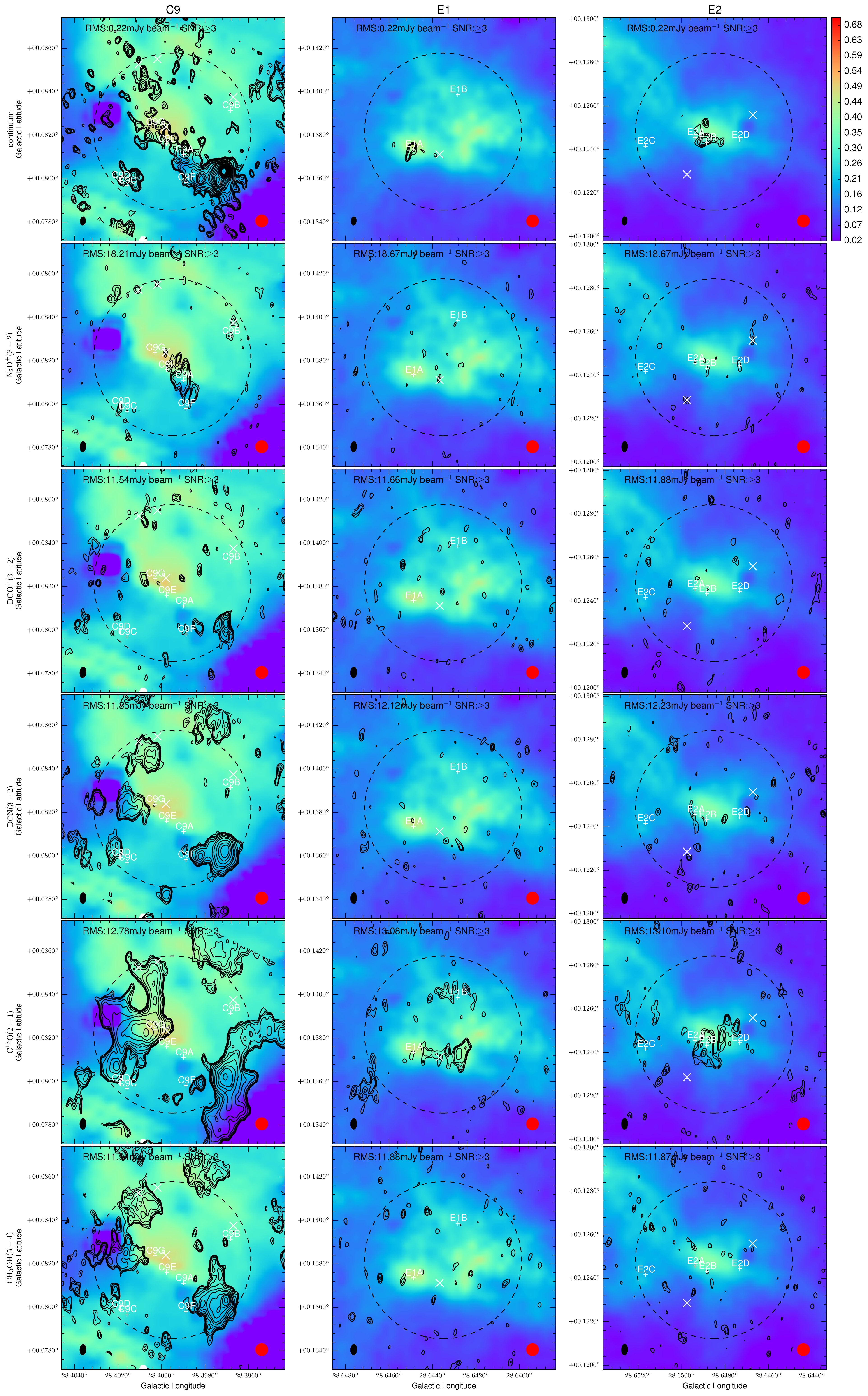}
\caption{
Same as Fig. \ref{fig:a1a2a3}, but for sources C9, E1, E2.
\label{fig:c9e1e2}}
\end{figure*}

\begin{figure*}[htb!]
\epsscale{0.95}
\plotone{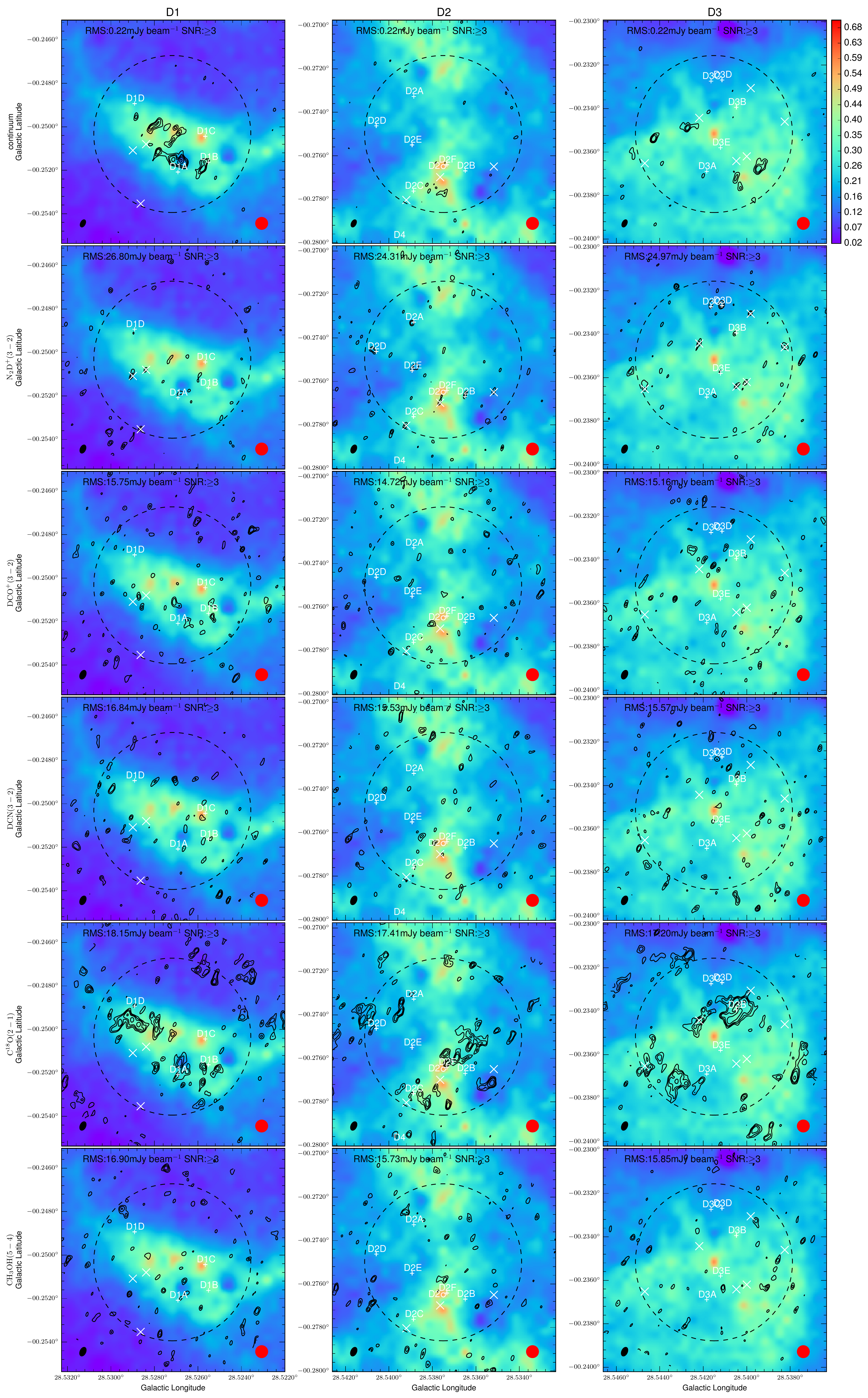}
\caption{
Same as Fig. \ref{fig:a1a2a3}, but for sources D1, D2, D3. Note, the
center of the D2 core/clump from BT12 is shown with an open
circle. The D4 complex is at the top of the D2 box, i.e., only partially
within the 26\arcsec\ diameter primary beam FOV, and the center of the
D4 core/clump from BT12 is just outside the displayed box. For naming
purposes, we assign all identified cores to D2.
\label{fig:d1d2d3}}
\end{figure*}

\begin{figure*}[htb!]
\epsscale{0.95}
\plotone{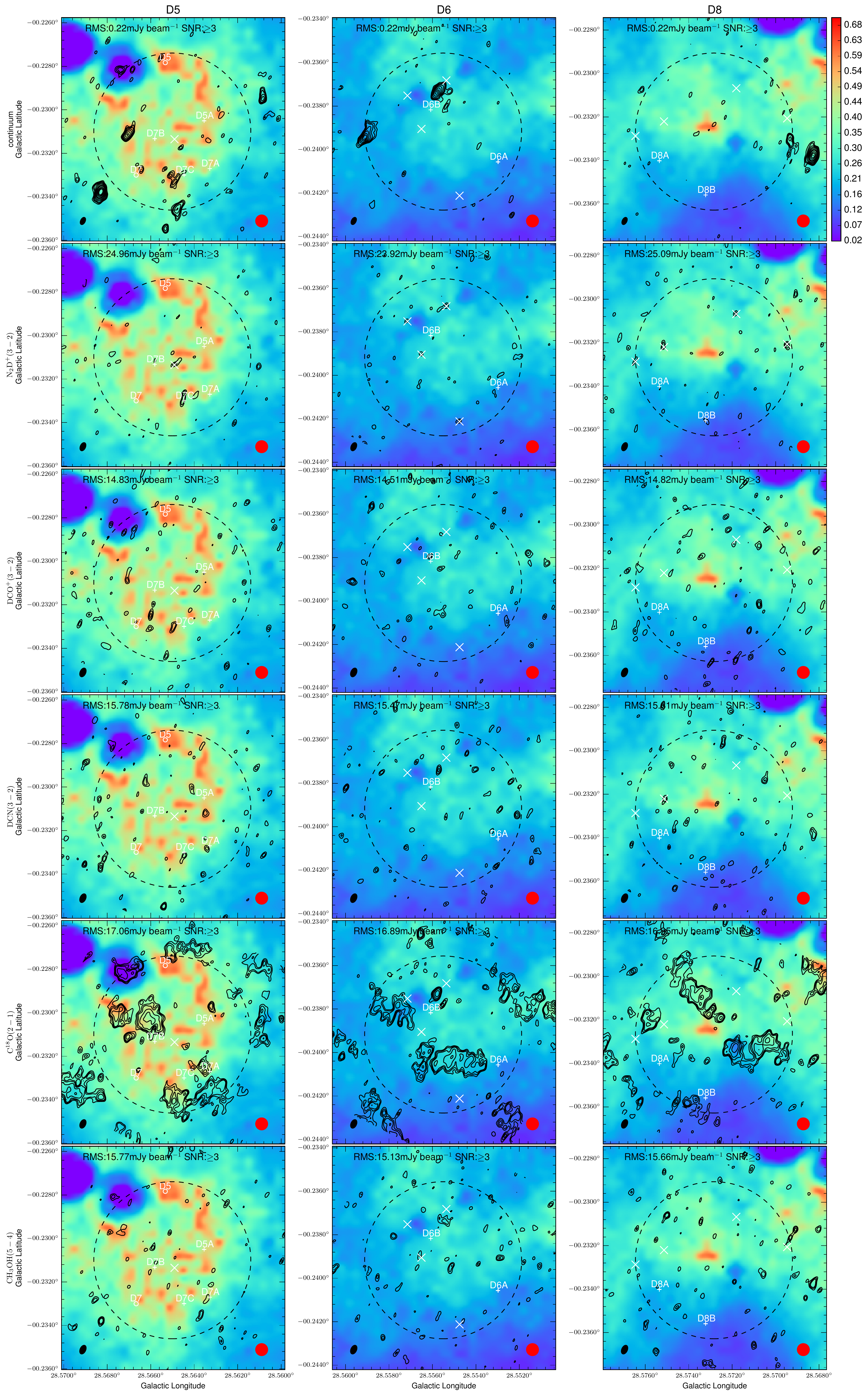}
\caption{
Same as Fig. \ref{fig:a1a2a3}, but for sources D5, D6, D8. Note, the
D5 box also contains D7 and the centers of these core/clumps from BT12
are shown with open circles.
\label{fig:d5d6d8}}
\end{figure*}

\begin{figure*}[htb!]
\epsscale{0.95}
\plotone{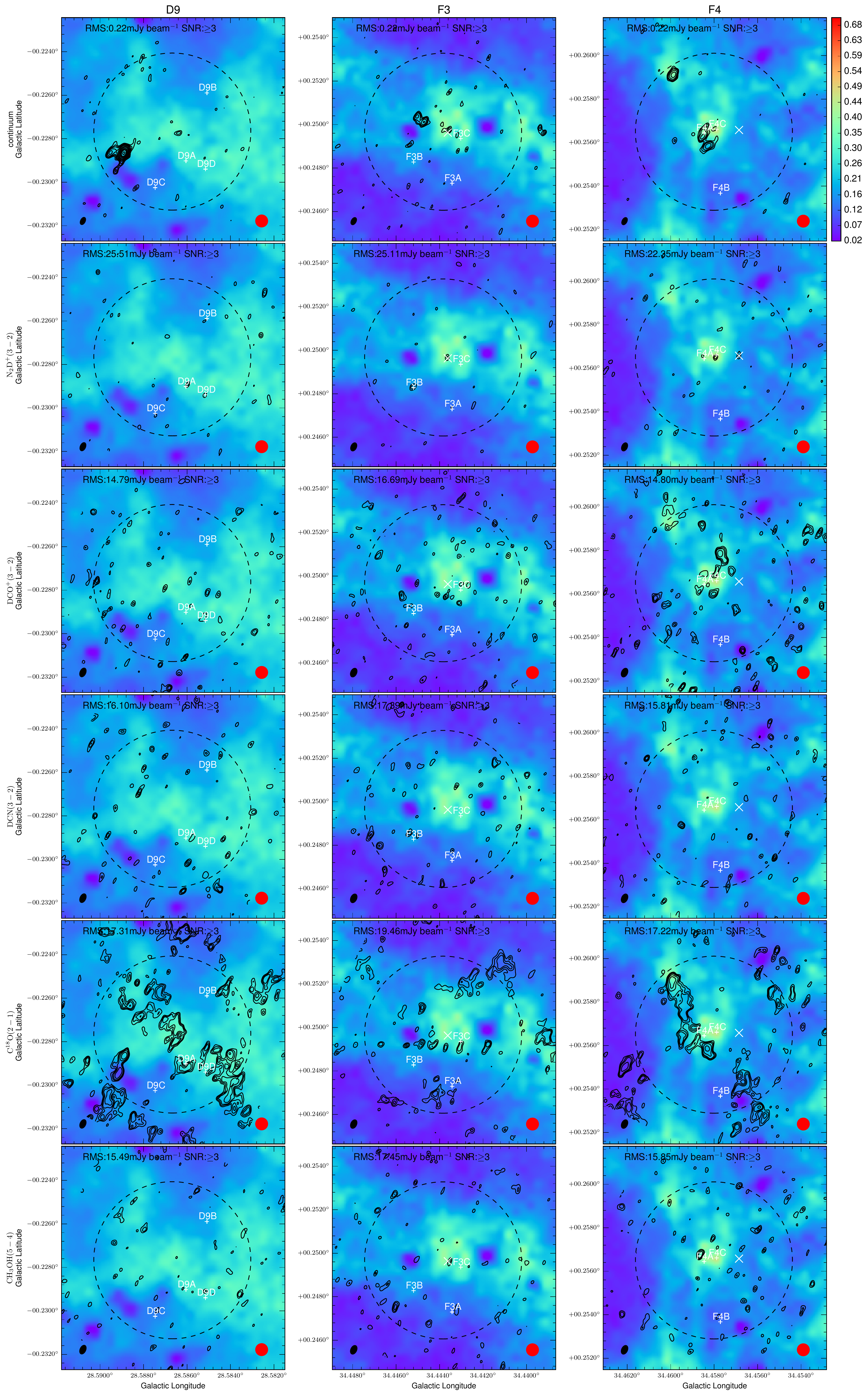}
\caption{
Same as Fig. \ref{fig:a1a2a3}, but for sources D9, F3, F4.
\label{fig:d9f3f4}}
\end{figure*}

\begin{figure*}[htb!]
\epsscale{0.95}
\plotone{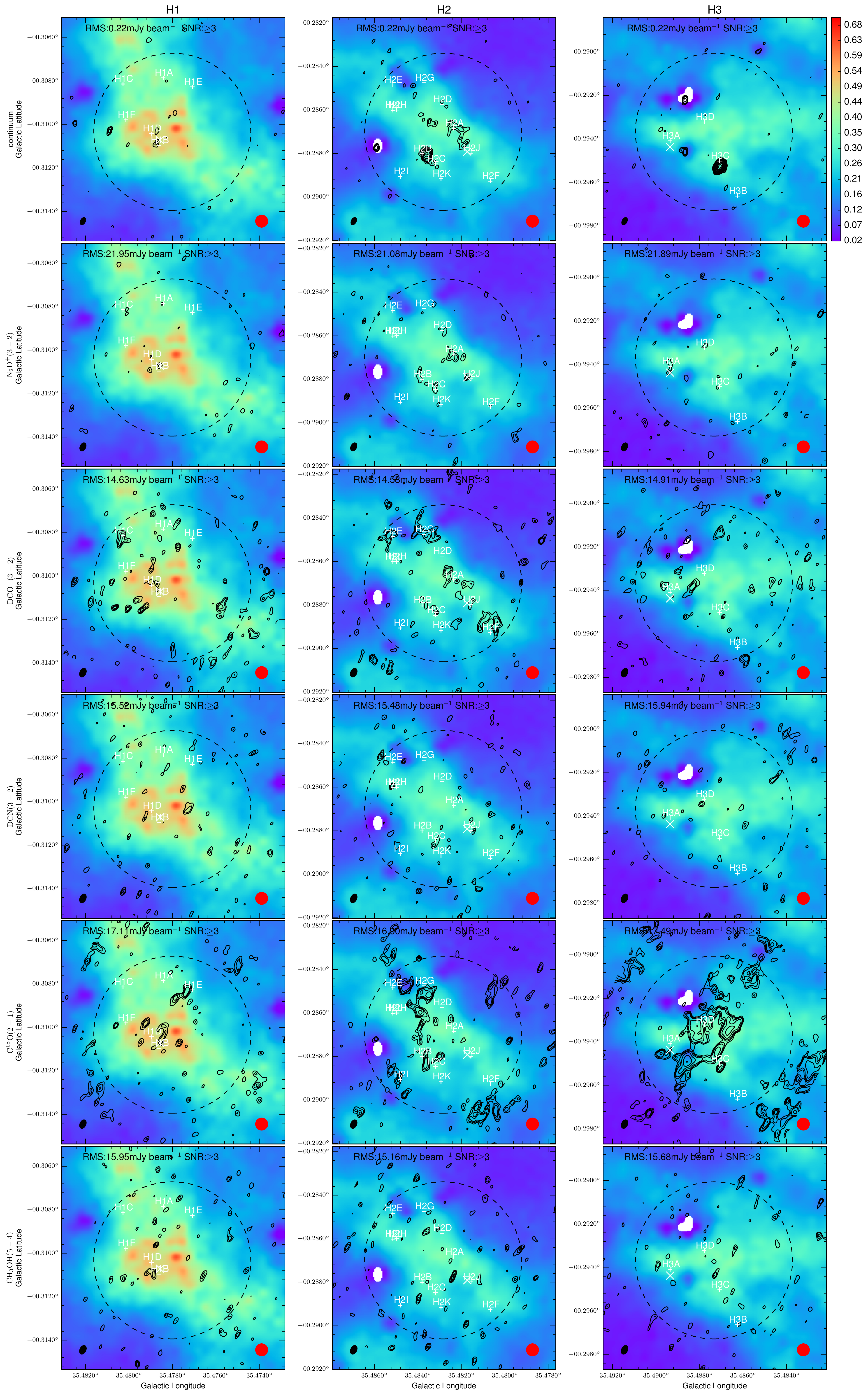}
\caption{
Same as Fig. \ref{fig:a1a2a3}, but for sources H1, H2, H3.
\label{fig:h1h2h3}}
\end{figure*}

\begin{figure*}[htb!]
\epsscale{0.95}
\plotone{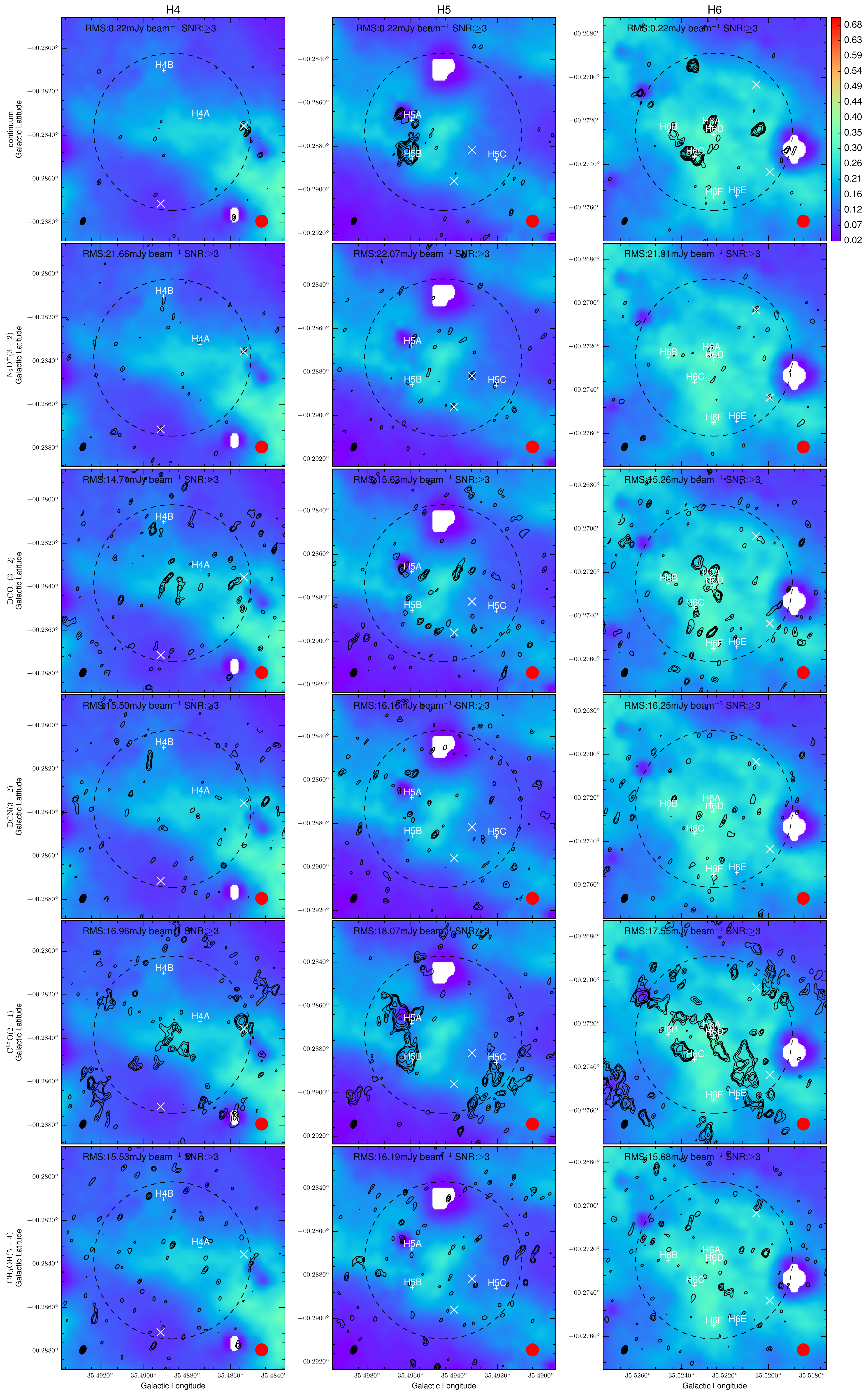}
\caption{
Same as Fig. \ref{fig:a1a2a3}, but for sources H4, H5, H6.
\label{fig:h4h5h6}}
\end{figure*}

Figures \ref{fig:a1a2a3} through \ref{fig:h4h5h6} present summary maps
of the 30 targets, one column for each source. Note, except in the
cases of D2 and D5, the FOV is centered on the IRDC core/clump
$\Sigma$ peak from BT12. The D2 region includes both D2 and (part of)
D4, while the D5 region includes both D5 and D7. In these cases the
FOV is centered in between the $\Sigma$ peaks. In all panels the
colored background shows MIREX-estimated mass surface density,
$\Sigma$, in $\rm g\:cm^{-2}$ from BT12. All panels also show the
primary beam diameter FOV, which delimits the region that is searched
for \ntdpns(3-2) cores. The locations (``$+$'' signs) and names (XXA,
XXB, XXC... indicates the ordering of decreasing \ntdpns(3-2) line
flux) of these cores are also shown. These cores are discussed below
in \S\ref{S:cores}. The panels also indicate the locations
(``$\times$'' signs) of \ntdpns(3-2) ``clump-scale'' structures with
$\geq4\sigma$ peaks in the 5~\kms velocity range 0th-moment map that
are not found by our \ntdpns(3-2) core detection algorithm.
 
In the top row the contours show 1.3~mm continuum emission. All
regions have generally very strong continuum detection (C3 only has a
3$\sigma$ detection, while D2 has 4$\sigma$ detection). There is often
quite good correspondence of the 1.3~mm continuum structures with
those seen in the BT12 MIREX maps. Note, however, that the mm continuum
image is insensitive to the structures that are larger than the
  maximum recoverable scale of the observation (i.e., $\gtrsim$
  20\arcsec), unlike the MIREX map. Still, there are many regions
where a MIREX $\Sigma$ peak is not an especially strong mm continuum
source and vice versa. Such discrepancies may arise because of
problems in the MIREX map (e.g., where MIR bright sources are present)
and/or if the mm continuum emission is being enhanced by higher
temperatures, e.g., from local protostellar heating. A lack of
prominent mm continuum emission from high $\Sigma$ MIREX peaks may
indicate these regions are extremely cold.

In the 2nd row the contours show \ntdpns(3-2) integrated intensity
(0th-moment map). Here, and for all 0th-moment maps, the integration
spans a velocity range of 5~\kms centered on $v_{\rm LSR}$ of each
IRDC\footnote{ 
The choice of a 5~\kms integration range makes this analysis
consistent with that of T13, which itself was motivated by our
previous single dish studies of these IRDCs.  The line-width of
\ntdpns(3-2) is $\sim$ 1 \kmsns, so a 5 \kms integration should be
enough to include all the cores in these regions, while at the same
time limiting the amount of extraneous noise. We have carried out
tests with 3, 4, 5 and 6 \kms integrations and find comparable
results. If a 4$\sigma$ contour shows up in the 5 \kms integration
(fiducial), there are signals at the same position in other maps.}  (see
Table~\ref{tab:obs}). We find the strongest detections of \ntdpns(3-2)
in the B1, C9 and H2 clumps, which all show extended structures that
are also identified as some of the highest \ntdpns(3-2) line flux
cores. In general, the strongest cores in a given region (e.g., A1A,
A3A, etc) are easily visible in these 0th-moment maps, but many of the
weaker cores do not stand out above the noise (lowest contour is
3$\sigma\simeq60$~mJy~bm$^{-1}$~\kmsns), which is higher than the
noise resulting from the more localized velocity ranges used to detect
the cores.

In addition there remain some $\geq4\sigma$ features visible in these
region maps that have not been identified as cores (i.e., as connected
structures in PPV space). Note, however that our core search is
confined to within the primary beam so strong features beyond this
scale would not have been identified. Since our focus is on the cores,
we simply note here the position of these ``clump-scale'' structures
(on average there are about 2 per region) and discuss below their
likelihood of being real features or simply noise fluctuations by
comparison with other tracers.

The 3rd to 6th rows show 0th-moments maps of \dcopns(3-2), DCN(3-2),
\ceions(2-1) and CH$_3$OH($v$ $t=0$ 5(1,4)-4(2,2)),
respectively. These images contain a lot of ancillary information,
which will not be discussed in detail here in this paper. We note that
all regions have strong \ceions(2-1) detection, which tends to trace
spatially extended structures (again subject to the constraints of
spatial filtering of the largest scales). The other tracers tend to
identify smaller scale structures.

One particular use of these images is to show the presence or absence
of these species at the locations of identified \ntdpns(3-2) cores and
``clump-scale'' structures. Cores will be discussed in more detail
below in \S\ref{S:cores}. For the clump-scale structures, several are
also seen in these ancillary tracers, e.g., in: DCO$^+$(3-2), such as
the source in A3; DCN(3-2), such as the MIR-bright source in B2; or
1.3~mm continuum, such as sources in C6 (north) and F3 (also seen in
CH$_3$OH). When such coincident detections in independent tracers are
seen, it increases our confidence in the reality of the \ntdpns(3-2)
detected structure. For the remainder it is hard to be sure about
their reality: more sensitive follow-up observations are needed. There
may be several reasons these \ntdpns(3-2) structures are not
identified as cores: they may not be well connected in velocity space;
they may not subtend a sufficient angular area compared to the beam;
they may be too close to the boundary of the primary beam, where we
truncate the core search algorithm; they may already be attached to an
idenified core, but have their clump-scale 0th-moment peak slightly
displaced from the core peak (e.g., near H2J and H3A).

Physically, some of the \ntdpns(3-2) structures that are only seen in
the broader, 5~\kms velocity range 0th-moment map may represent an
early stage of \ntdpns(3-2) core formation, i.e., when the
  material is less concentrated in position-velocity space. Thus the
follow-up of these structures with higher sensitivity observations is
warranted to investigate such a possibility.

\subsection{Identified \ntdp Cores}\label{S:cores}

\begin{deluxetable*}{cccccccccc}
\tablecolumns{10}
\tablewidth{0pc}
\tablecaption{\ntdpns(3-2) Cores\label{tab:results}}
\tablehead{
\colhead{Core} & \colhead{$l$} & \colhead{$b$} &\colhead{$v_{\rm min}$} & \colhead{$v_{\rm max}$} & \colhead{$\bar v$} &\colhead{$S_{\rm N_2D^+}$} & \colhead{$S_{\rm 1.30mm}$/SNR} & \colhead{$S_{\rm DCO^+}$/SNR} & \colhead{$S_{\rm C^{18}O}$/SNR}\\
\colhead{} & \colhead{(deg)} & \colhead{(deg)} &\colhead{(\kmsns)} & \colhead{(\kmsns)} & \colhead{(\kmsns)} &\colhead{(Jy~\kmsns)} & \colhead{\jyb\kmsns} & \colhead{\jyb\kmsns} & \colhead{\jyb\kmsns} \\
\colhead{(1)} & \colhead{(2)} & \colhead{(3)} &\colhead{(4)} & \colhead{(5)} & \colhead{(6)} &\colhead{(7)} & \colhead{(8)} & \colhead{(9)} & \colhead{(10)}
}
\startdata
1 C9A & 28.39894 & 0.08112 & 77.05 & 80.35 & 78.44 & 1.4e+00 & 6.9e-03/30 & -8.6e-03/-0.88 & -2.3e-01/-21 \\ 
C1S & 28.32194 & 0.06737 & 76.90 & 81.90 & 79.40 & 6.3e-01 & 1.6e-02/59 & 6.7e-02/8.2 & -/- \\ 
G2N & 34.78101 & -0.56816 & 38.95 & 43.95 & 41.45 & 6.1e-01 & 1.4e-03/8.3 & 1.5e-01/14 & -/- \\ 
2 B1A & 19.28744 & 0.08048 & 26.40 & 29.10 & 27.66 & 4.6e-01 & 3.4e-04/1.5 & 4.4e-02/5.0 & 2.6e-02/2.7 \\ 
C1N & 28.32508 & 0.06714 & 78.68 & 83.68 & 81.18 & 2.7e-01 & 1.9e-03/6.9 & 4.6e-02/5.6 & -/- \\ 
3 H2A & 35.48231 & -0.28684 & 44.95 & 47.20 & 45.60 & 2.3e-01 & 5.4e-04/2.4 & 4.7e-02/4.7 & 2.7e-02/2.3 \\ 
G2S & 34.77842 & -0.56838 & 39.30 & 44.30 & 41.80 & 2.2e-01 & 1.2e-03/6.7 & 1.1e-01/10 & -/- \\ 
F1 & 34.41928 & 0.24588 & 53.62 & 58.62 & 56.12 & 1.6e-01 & 2.9e-03/8.7 & 1.0e-01/12 & -/- \\ 
F2 & 34.43525 & 0.24140 & 55.16 & 60.16 & 57.66 & 9.5e-02 & 3.5e-03/16 & 1.5e-01/16 & -/- \\ 
4 B1B & 19.28583 & 0.08304 & 26.40 & 27.15 & 26.76 & 6.5e-02 & 7.6e-03/23 & 3.3e-02/7.0 & 6.7e-03/1.3 \\ 
5 C9B & 28.39679 & 0.08314 & 76.45 & 77.95 & 77.30 & 6.5e-02 & 1.1e-04/0.32 & -1.2e-02/-1.7 & -5.4e-02/-7.5 \\ 
6 H2B & 35.48377 & -0.28801 & 45.10 & 46.90 & 46.10 & 6.3e-02 & 2.6e-03/11 & 2.5e-02/2.7 & 4.1e-02/3.9 \\ 
7 D6A & 28.55294 & -0.24057 & 85.50 & 86.85 & 86.39 & 6.0e-02 & 3.5e-04/0.93 & -1.0e-02/-1.2 & 1.1e-02/1.2 \\ 
8 D5A & 28.56350 & -0.23050 & 87.90 & 88.95 & 88.28 & 5.8e-02 & 4.3e-04/1.7 & 2.6e-03/0.33 & -2.3e-02/-2.9 \\ 
9 B2A & 19.30983 & 0.06697 & 24.90 & 26.55 & 25.56 & 5.6e-02 & 2.8e-03/9.8 & 2.8e-02/4.2 & 5.5e-02/7.5 \\ 
10 H5A & 35.49591 & -0.28680 & 45.10 & 46.30 & 45.34 & 5.2e-02 & 4.1e-04/1.7 & 3.3e-02/3.9 & 2.0e-02/2.2 \\ 
11 H1A & 35.47838 & -0.30785 & 44.20 & 45.55 & 44.88 & 5.0e-02 & -2.2e-04/-0.68 & 2.5e-02/3.0 & 2.3e-02/2.5 \\ 
12 H2C & 35.48314 & -0.28847 & 44.95 & 46.45 & 45.69 & 4.9e-02 & 9.0e-04/3.6 & 2.1e-02/2.6 & 3.0e-02/3.1 \\ 
13 F4A & 34.45842 & 0.25645 & 56.95 & 58.15 & 57.40 & 4.9e-02 & 1.4e-03/6.4 & 4.7e-02/6.0 & 1.5e-02/1.7 \\ 
14 C3A & 28.34951 & 0.09583 & 77.35 & 79.00 & 78.29 & 4.9e-02 & 3.8e-04/1.1 & 6.7e-03/1.0 & -1.7e-02/-2.4 \\ 
15 C5A & 28.35517 & 0.05754 & 76.45 & 77.65 & 76.80 & 4.8e-02 & -7.0e-05/-0.29 & 4.0e-03/0.72 & 1.4e-03/0.22 \\ 
16 D2A & 28.53882 & -0.27327 & 84.45 & 85.80 & 84.73 & 4.8e-02 & 1.2e-04/0.40 & 4.0e-04/0.048 & -9.0e-03/-0.94 \\ 
17 H2D & 35.48285 & -0.28575 & 44.35 & 47.20 & 45.68 & 4.6e-02 & 2.8e-04/1.2 & 2.7e-02/2.4 & -5.6e-02/-4.3 \\ 
18 H2E & 35.48511 & -0.28485 & 45.55 & 46.60 & 46.08 & 4.6e-02 & 1.9e-04/0.51 & 5.4e-02/7.5 & 3.5e-04/0.044 \\ 
19 D2B & 28.53645 & -0.27670 & 85.50 & 86.55 & 86.11 & 4.6e-02 & 1.3e-05/0.047 & -1.0e-02/-1.3 & -2.5e-03/-0.31 \\ 
20 C4A & 28.35450 & 0.07184 & 79.15 & 80.65 & 79.87 & 4.4e-02 & 5.0e-04/2.2 & 8.5e-03/1.4 & 2.3e-02/3.3 \\ 
21 C2A & 28.34592 & 0.05881 & 77.35 & 79.75 & 78.36 & 4.3e-02 & 5.9e-04/1.9 & 8.1e-03/0.94 & -1.6e-02/-1.7 \\ 
22 C2B & 28.34677 & 0.05918 & 79.45 & 79.60 & 79.50 & 4.1e-02 & 2.4e-03/6.5 & -3.1e-03/-0.92 & -4.9e-03/-1.8 \\ 
23 A3A & 18.80730 & -0.30530 & 64.10 & 66.05 & 65.16 & 3.9e-02 & 1.8e-03/8.2 & 2.4e-02/3.3 & 1.8e-02/2.4 \\ 
24 D1A & 28.52688 & -0.25208 & 87.30 & 87.90 & 87.46 & 3.8e-02 & 1.3e-04/0.50 & 1.1e-02/1.8 & 3.0e-02/4.5 \\ 
25 A3B & 18.80692 & -0.30445 & 63.95 & 65.90 & 65.15 & 3.7e-02 & 4.3e-05/0.18 & 1.5e-02/2.1 & 2.3e-02/3.1 \\ 
26 H1B & 35.47858 & -0.31096 & 43.90 & 45.40 & 44.49 & 3.6e-02 & 5.5e-04/2.4 & 9.3e-03/1.1 & -8.1e-03/-0.84 \\ 
27 D2C & 28.53883 & -0.27762 & 88.65 & 88.80 & 88.75 & 3.5e-02 & 1.9e-04/0.57 & -3.3e-03/-0.79 & 4.3e-03/1.0 \\ 
28 D1B & 28.52548 & -0.25163 & 86.40 & 86.70 & 86.58 & 3.5e-02 & 6.6e-04/2.4 & 3.3e-03/0.74 & 1.1e-05/2.1e-3 \\ 
29 C7A & 28.36254 & 0.12185 & 79.45 & 80.50 & 80.20 & 3.4e-02 & 2.8e-04/0.76 & -2.5e-03/-0.42 & -2.7e-02/-4.6 \\ 
30 D7A & 28.56324 & -0.23270 & 84.45 & 84.90 & 84.65 & 3.3e-02 & 1.2e-04/0.40 & -2.4e-03/-0.46 & -1.1e-02/-1.8 \\ 
31 D2D & 28.54054 & -0.27463 & 88.80 & 89.25 & 89.09 & 3.3e-02 & -1.0e-04/-0.29 & 6.1e-04/0.12 & -2.6e-03/-0.48 \\ 
32 D8A & 28.57531 & -0.23401 & 84.45 & 85.50 & 84.65 & 3.3e-02 & -2.5e-04/-0.78 & 1.3e-02/1.8 & -2.7e-03/-0.32 \\ 
33 D3A & 28.54180 & -0.23689 & 85.80 & 86.85 & 86.66 & 3.3e-02 & 2.8e-04/1.1 & -3.3e-02/-4.5 & 1.6e-02/2.0 \\ 
34 C4B & 28.35479 & 0.07069 & 80.05 & 80.20 & 80.14 & 3.2e-02 & 9.8e-04/4.4 & 2.7e-02/8.6 & 8.2e-03/3.3 \\ 
35 D9A & 28.58601 & -0.22902 & 85.95 & 87.45 & 86.63 & 3.2e-02 & -8.4e-05/-0.34 & 2.1e-03/0.25 & -4.8e-04/-0.049 \\ 
36 D2E & 28.53890 & -0.27550 & 87.15 & 88.65 & 88.01 & 3.0e-02 & -1.7e-04/-0.71 & -5.9e-03/-0.71 & -3.1e-02/-3.1 \\ 
37 B2B & 19.30633 & 0.06616 & 25.95 & 27.45 & 26.74 & 2.9e-02 & 1.2e-03/5.2 & 3.0e-02/4.4 & 3.1e-02/4.4 \\ 
38 C9C & 28.40155 & 0.07969 & 78.70 & 79.30 & 78.97 & 2.8e-02 & -8.8e-04/-2.2 & 4.2e-03/0.92 & -5.2e-03/-1.1 \\ 
39 C4C & 28.35621 & 0.07074 & 79.15 & 80.95 & 79.95 & 2.8e-02 & -2.0e-05/-0.074 & 5.1e-03/0.72 & 1.2e-02/1.6 \\ 
40 H6A & 35.52262 & -0.27226 & 44.95 & 46.15 & 45.49 & 2.7e-02 & 4.8e-03/22 & 5.0e-02/6.1 & 1.6e-01/18 \\ 
41 D1C & 28.52562 & -0.25044 & 86.55 & 86.70 & 86.64 & 2.6e-02 & 3.0e-05/0.12 & 1.1e-03/0.36 & -5.4e-03/-1.2 \\ 
42 C2C & 28.34311 & 0.06017 & 78.40 & 80.20 & 79.14 & 2.6e-02 & 8.8e-03/39 & 2.9e-02/3.8 & 1.3e-01/15 \\ 
43 H2F & 35.48063 & -0.28927 & 45.10 & 46.60 & 46.02 & 2.5e-02 & 5.2e-04/1.4 & -5.6e-04/-0.068 & -2.4e-02/-2.5 \\ 
44 H3A & 35.48931 & -0.29409 & 42.40 & 44.65 & 43.33 & 2.5e-02 & 2.9e-04/1.1 & 4.0e-02/3.9 & -3.4e-03/-0.28 \\ 
45 D7B & 28.56577 & -0.23133 & 89.10 & 89.25 & 89.24 & 2.5e-02 & -1.4e-04/-0.62 & 2.7e-03/0.65 & -7.8e-03/-2.3 \\ 
46 C6A & 28.36492 & 0.05084 & 80.20 & 81.25 & 80.63 & 2.4e-02 & -3.2e-04/-1.1 & 1.2e-02/2.3 & 8.0e-03/1.4 \\ 
47 H2G & 35.48368 & -0.28475 & 45.10 & 45.25 & 45.25 & 2.4e-02 & 3.2e-05/0.11 & 1.7e-02/5.9 & 6.4e-03/1.5 \\ 
48 D3B & 28.54043 & -0.23394 & 87.60 & 88.65 & 87.83 & 2.4e-02 & 4.6e-04/1.9 & 1.5e-02/2.0 & 3.8e-02/4.7 \\ 
49 D8B & 28.57318 & -0.23559 & 88.50 & 89.25 & 88.72 & 2.3e-02 & 1.2e-04/0.35 & -2.4e-03/-0.36 & -8.2e-03/-1.2 \\ 
50 D2F & 28.53729 & -0.27641 & 87.30 & 87.45 & 87.40 & 2.3e-02 & 7.2e-04/3.0 & -2.8e-03/-0.67 & -4.0e-03/-1.2  
\enddata
\end{deluxetable*}


In total 141 \ntdpns(3-2) cores are found in the 30 data cubes, based
on the criteria stated in \S\ref{sec:method}. They are labeled as $+$
signs in the 2nd row of Figures~\ref{fig:a1a2a3} through
\ref{fig:h4h5h6}. The cores are named based on the ranking of their
\ntdpns(3-2) flux within each surveyed region: e.g., A3A is the
highest flux core within the A3 region, followed by A3B, A3C, etc.

Most strong features in the 0th-moment maps are identified, along with
many cores that do not show up in these maps. Note, these 0th-moment
maps are integrated over a 5~\kms range. Meanwhile, many weak \ntdp
cores 
are typically found in a velocity range much less than 5~\kms (the
narrowest velocity range is 0.30~\kms by definition, corresponding to
two channels, see \S\ref{sec:method}).  As a result, many of the weak
cores do not show up in the region figures, i.e., at the positions of
these cores there is no corresponding \ntdp contours seen.  


We have ranked the cores based on their \ntdp 0th-moment flux, and
list the strongest 50 of these in Table \ref{tab:results}, together
with the 6 cores from T13.
The table lists core positions (intensity-weighted center) in Galactic
coordinates in columns 2 and 3. Columns 4 and 5 give the velocity
range over which the core is detected. Note the \ntdpns(3-2) cube used
for core finding has been smoothed to 0.15~\kms velocity resolution.
In general, cores with structures in more channels tend to be more
significant. Column 6 shows the intensity weighted $v_{\rm
  LSR}$. Column 7 shows the \ntdpns(3-2) total line flux, $S_{\rm
  N_2D^+}$, for each core. Columns 8, 9 and 10 show fluxes at the core
position in the continuum image, the \dcop(3-2) 0th-moment image, and
the \ceio(2-1) 0th-moment image (note the T13 spectral set-up did not
have \ceio(2-1)).  The 0th-moment images of \dcop and \ceio are
integrated within the velocity ranges shown in columns 4 and 5. The
relevant signal-to-noise ratio (SNR) is listed in these columns also. Note, that sometimes
these fluxes can be negative in the case of non-detections where noise
or other fluctuations cause the signal to be negative.

\begin{figure*}[htb!]
\epsscale{1.}
\plotone{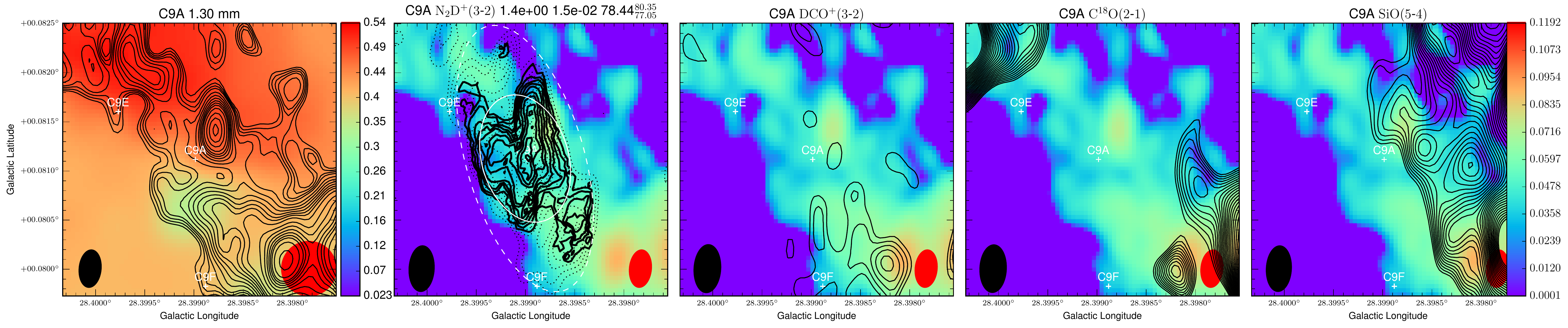}\\
\plotone{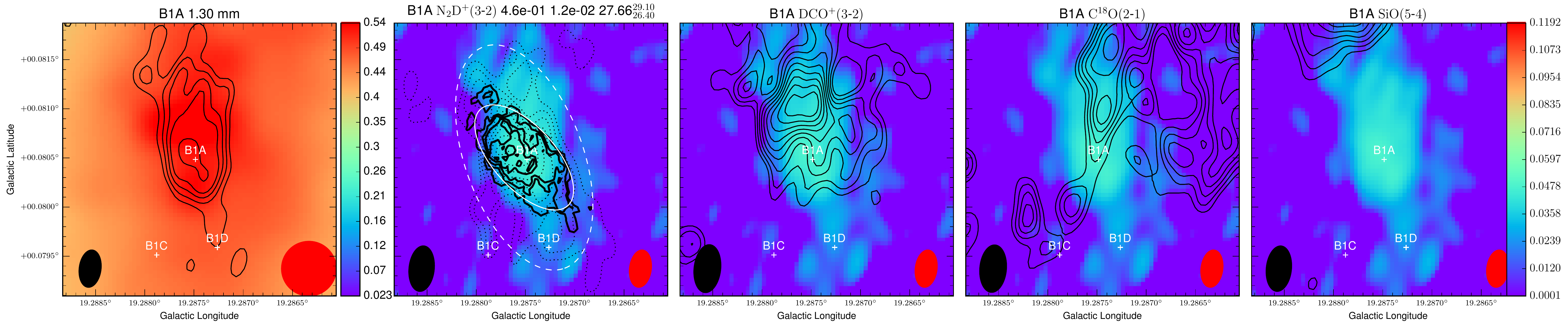}\\
\plotone{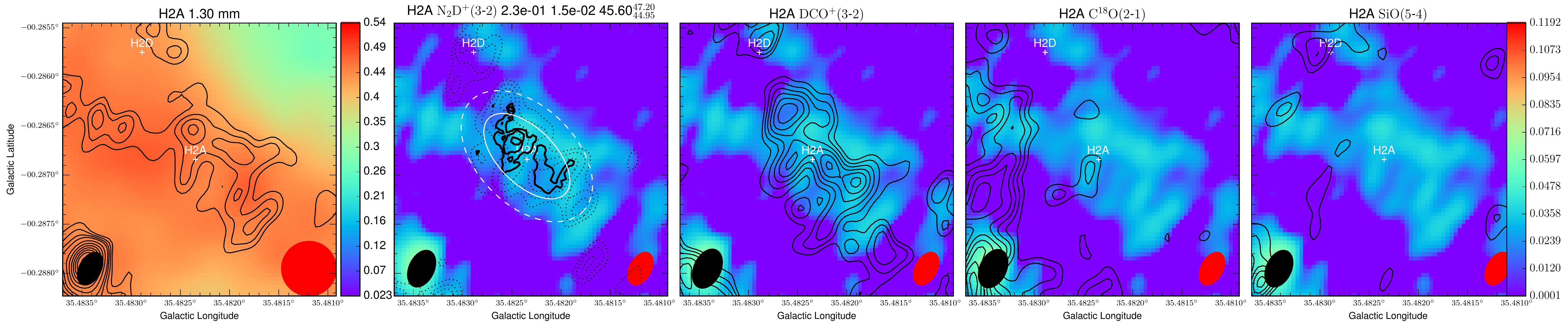}\\
\plotone{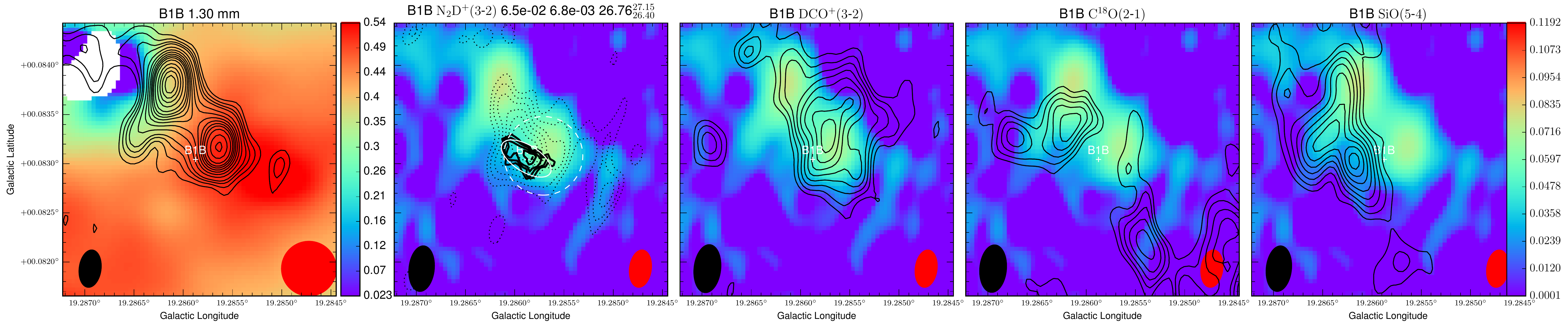}\\
\plotone{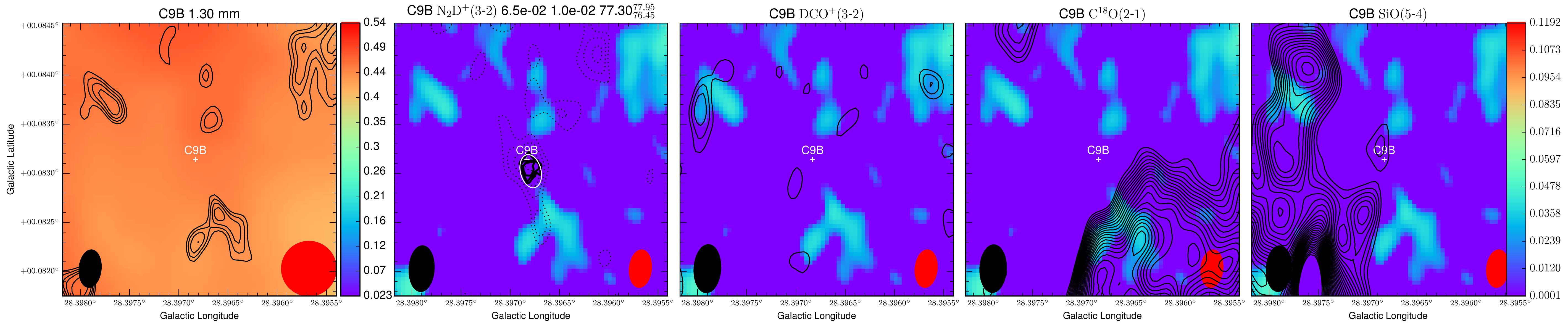}
\caption{
Core-scale zoom-in figures of C9A, B1A, H2A, B1B, C9B (rows top to
bottom, also following rank order of \ntdpns(3-2) flux). Each panel
shows a 10\arcsec\ by 10\arcsec\ FOV. The first column shows 1.3~mm
continuum (contours: 2, 3, 4, 5, 7, 9, 11, 13, 15, 17, 20, 40, 60,
80... $\sigma$, where $1\sigma \simeq 2.2\times 10^{-4}$~\jybns; {\it
  ALMA} beam in lower left) overlaid on MIREX (BT12) mass surface
density (color scale in g~cm$^{-2}$; {\it Spitzer} beam in lower
right).
2nd column shows 0th-moment \ntdpns(3-2) of the cores (only
including core voxels), shown as the solid contours: 2, 3, 4,
5...$\sigma$ (beam lower left), overlaid on the 1.30~mm continuum
(beam lower right). The dotted contours are the full integration of
the cores (including all voxels within the channels).  At the top of
the panel we list total \ntdpns(3-2) flux (in \jyb\kmsns),
\ntdpns(3-2) 0th-moment map RMS (in \jyb\kmsns), and \ntdpns(3-2)
intensity weighted core velocity and velocity range (in \kmsns). The
solid and dashed ellipses are fitted boundaries to the cores (see
text).  
3rd column is the same as the 2nd, except contours now show
\dcopns(3-2) integrated over the same velocity range as the \ntdp
core.
4th column is the same as the 2nd, except contours now show
\ceions(2-1) integrated over the same velocity range as the \ntdp
core.
5th column is the same as the 2nd, except contours now show SiO(5-4)
integrated over the 5~\kms velocity range used in the region maps.
\label{fig:rowfigstart}}
\end{figure*}

\begin{figure*}[htb!]
\epsscale{1.}
\plotone{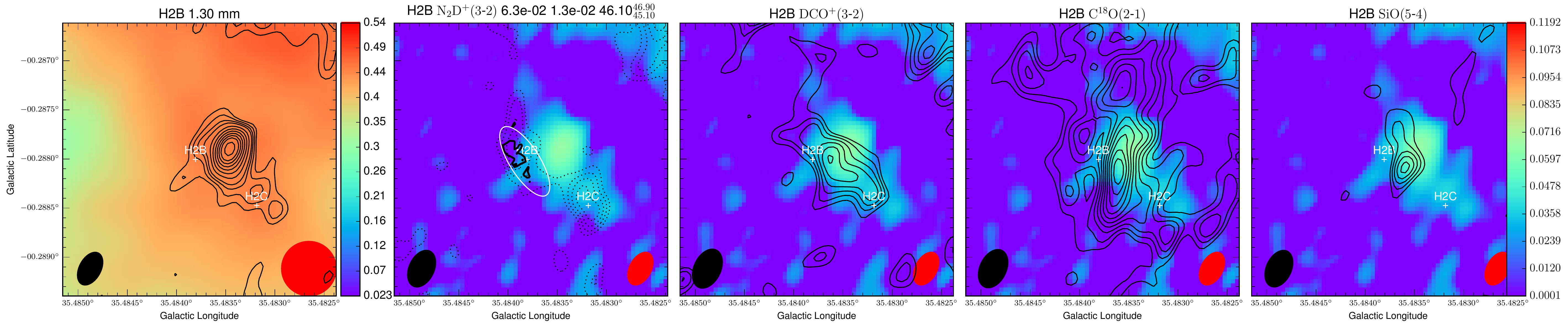}\\
\plotone{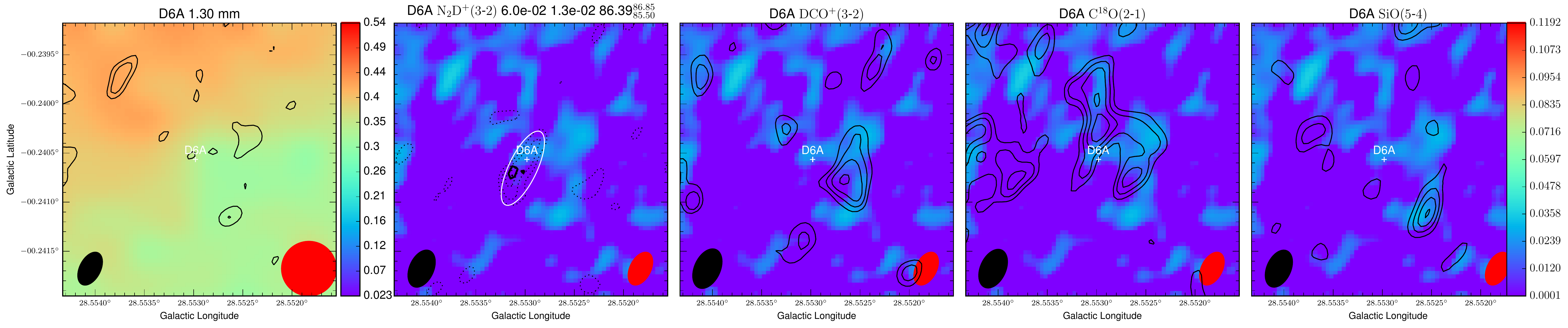}\\
\plotone{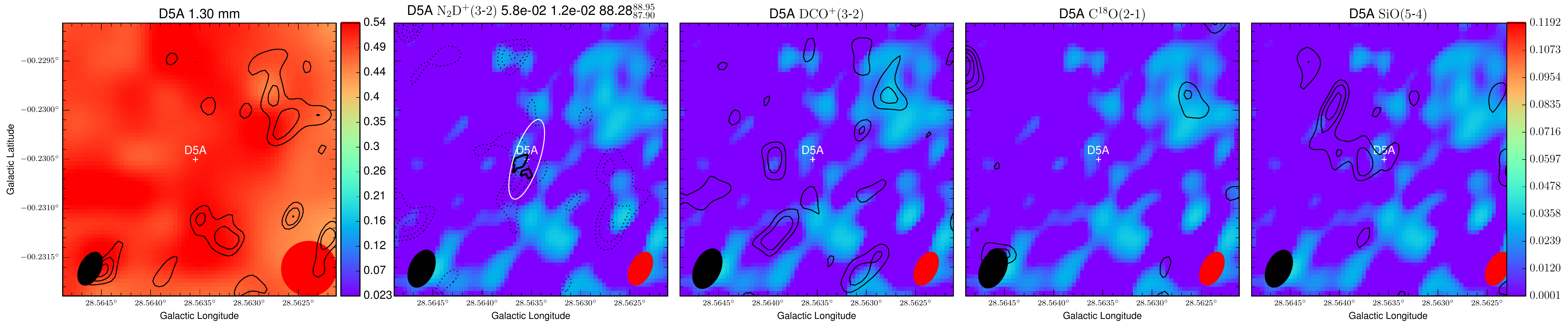}\\
\plotone{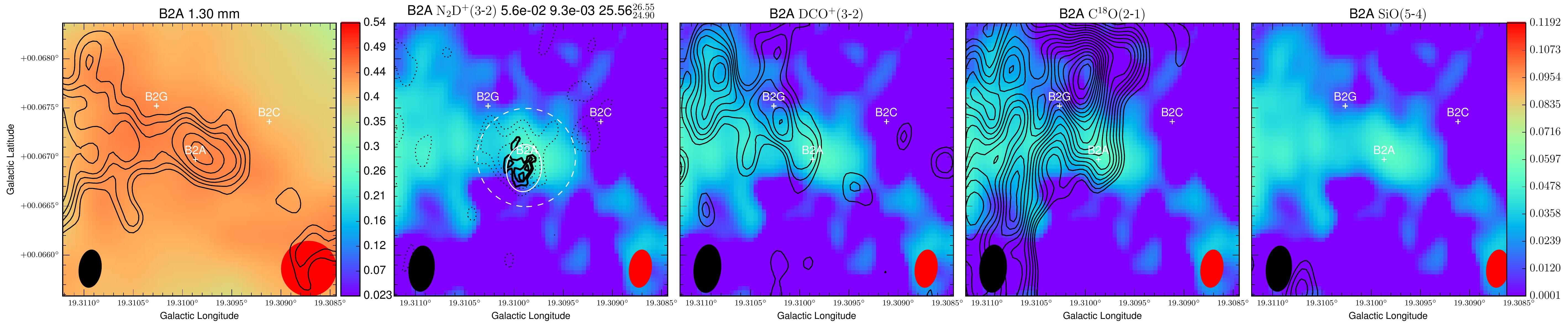}\\
\plotone{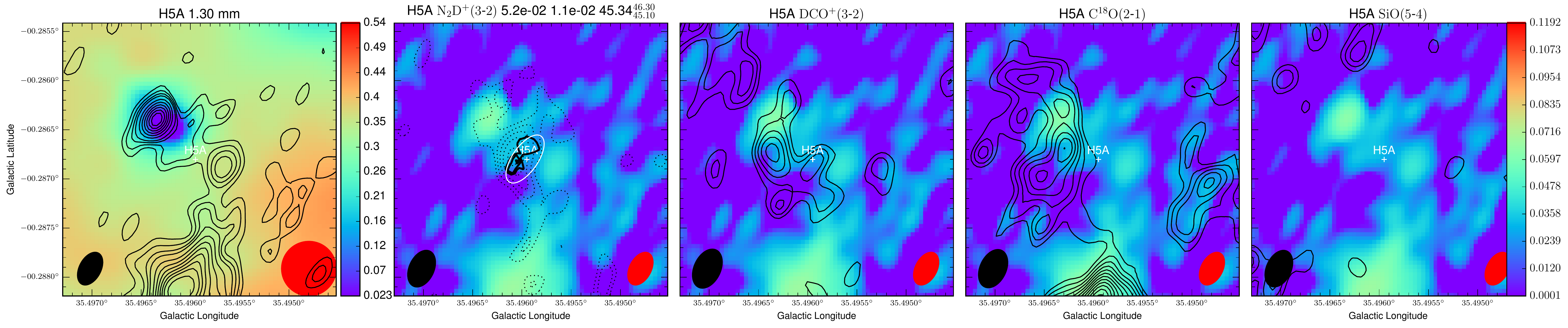}
\caption{
Same as Fig. \ref{fig:rowfigstart}, but for H2B, D6A, D5A, B2A, H5A. 
\label{fig:H2coreB}}
\end{figure*}

\begin{figure*}[htb!]
\epsscale{1.}
\plotone{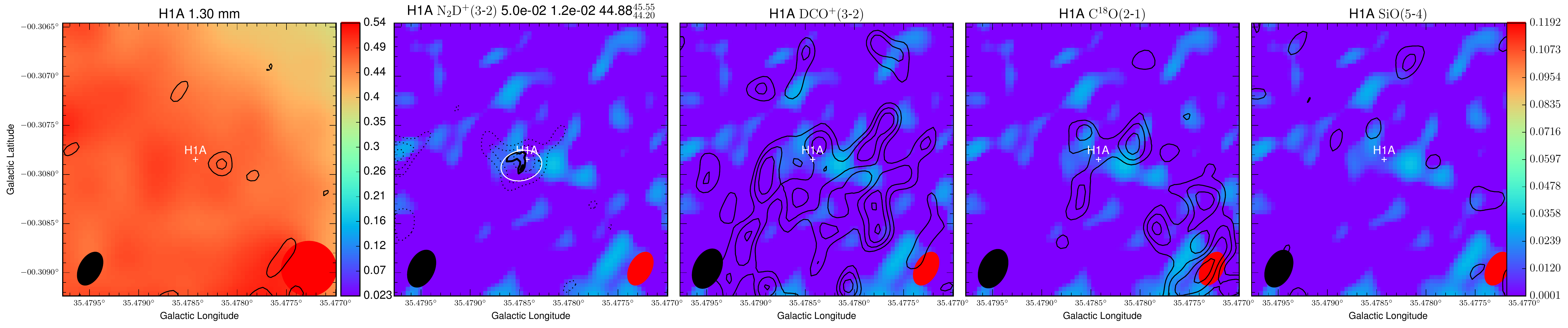}\\
\plotone{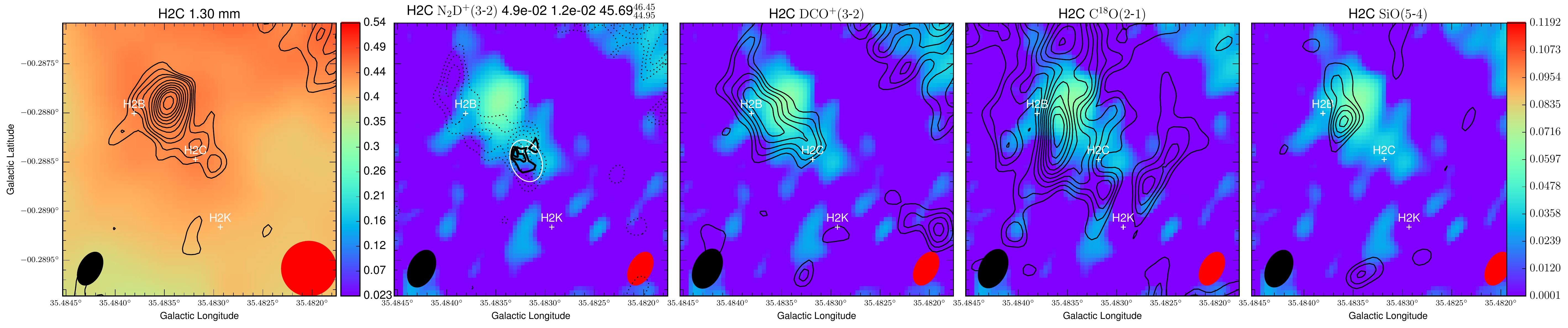}\\
\plotone{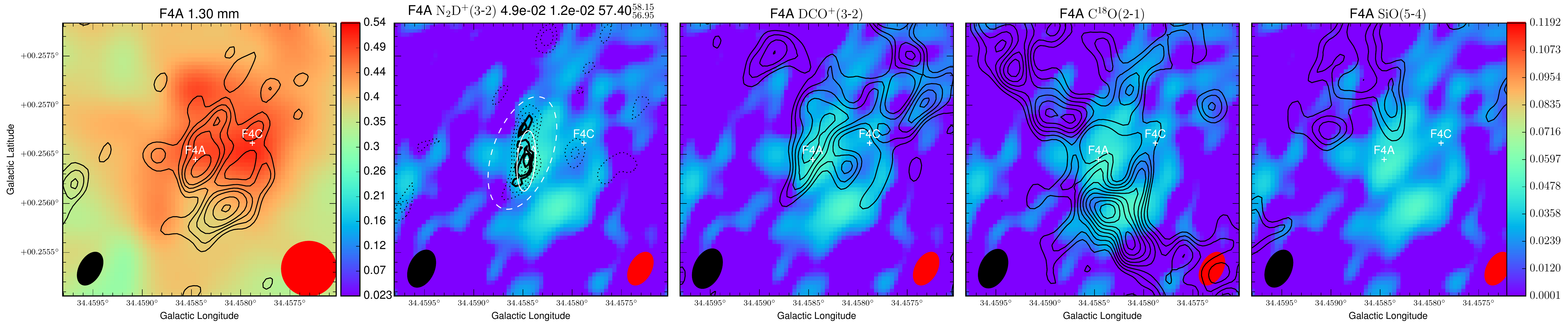}\\
\plotone{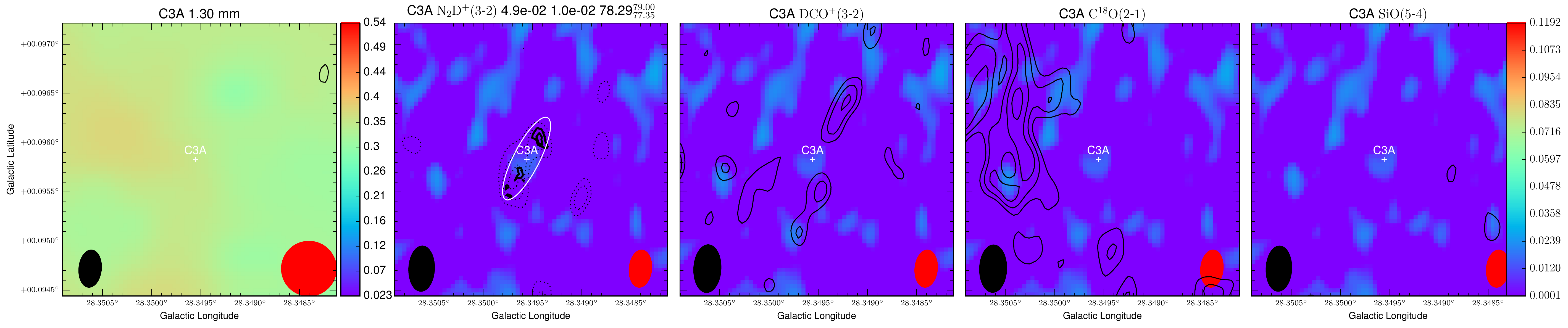}\\
\plotone{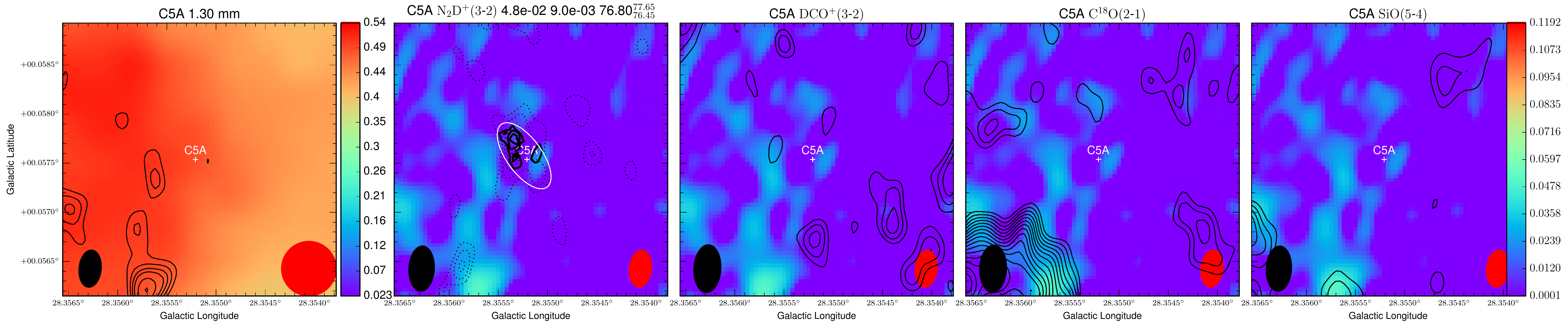}
\caption{
Same as Fig. \ref{fig:rowfigstart}, but for H1A, H2C, F4A, C3A, C5A. 
\label{fig:rowfigend}}
\end{figure*}

Figures \ref{fig:rowfigstart} to \ref{fig:rowfigend} show the top 15
newly detected \ntdpns(3-2) cores, ordered by \ntdpns(3-2) line flux.
Each figure row shows a 10\arcsec\ by 10\arcsec\ zoom-in view of the
core, centering at the \ntdpns(3-2) intensity weighted center.  The
first panel in each row shows the 1.3~mm continuum contours overlaid
on the MIREX map (BT12). For all figures we use the same color scale
for the MIREX image.

The second panel shows the \ntdp(3-2) integrated intensity (black
solid contours) on top of the 1.3~mm continuum emission (same color
scale for all sources).  The integration for these \ntdp(3-2)
0th-moment maps is only over velocity channels in which the core is
defined and only including voxels flagged to be in the core. We also
show a full 0th-moment integration map as grey dotted contours, which
includes all voxels within the velocity ranges. Both types of contour
maps start at 2$\sigma$ and increase with a step of 1$\sigma$, with
$\sigma$ being the noise of the full integration map. The grey dotted
contours always extend over a wider area than the black ones because
the full integration picks up more flux at the core boundary, which is
also why the black solid contours are sometimes separated. At the top
of the panel, several properties of the core: \ntdpns(3-2) flux,
\ntdpns(3-2) 0th-moment map RMS, and \ntdpns(3-2) intensity weighted
core velocity (sub- and super-scripts are velocity boundaries over
which core is detected).

The third, fourth and fifth panels show \dcopns(3-2), \ceio(2-1) and
SiO(5-4) 0th-moment maps on top of the dust continuum. For
\dcopns(3-2) and \ceio(2-1), the integrations are over the same
velocity ranges as the \ntdpns(3-2) 0th-moment map. For SiO(5-4) the
velocity range is the 5~\kms range used in the region maps (note a
full analysis of the SiO(5-4) data from the region maps will be
presented in a companion paper by Liu et al., in prep.). 


Comparing some of the top ranked cores, including those of T13 (see
Tabel~\ref{tab:results}), C9A has more than twice the \ntdpns(3-2) flux
of C1-S. B1A ranks above C1-N, and H2A ranks above F1. Note that the
integration for C9A, B1B, H2A only includes voxels in the connected
components (the defined cores), so their fluxes are higher in a full
integration (i.e., grey dotted contours in these core figures). However, we
also note that the sensitivity in this dataset is about 3 times worse
than T13, so the uncertainty in core fluxes is also higher.


One of the features of the 6 \ntdpns(3-2) cores found by T13 was their
extended \dcopns(3-2) envelopes. This is still true for most of the
new cores presented here, however, now we also see some cores, like
C9A, C9B, D5A, C3A and C5A, which have relatively weak \dcopns(3-2)
emission. This chemical diversity may reflect different environmental
conditions amongst the core sample.

We note that there is often a dearth of \ceions(2-1) emission
associated with the \ntdpns(3-2) cores, which likely indicates that
there is a high degree of gas phase depletion of CO, i.e., due to
freeze out onto dust grain ice mantles.  Such CO depletion is thought
to boost the deuteration of \nthp (e.g., K15), and so
this anti-correlation of CO emission with \ntdp emission is expected
theoretically. 



SiO(5-4) emission is a known tracer of protostellar outflows and so
can help us assess such activity within the cores. C9A and C9B show
extended SiO emission in their vicinity, which is however thought to
mostly arise from a separate massive protostellar source (e.g., that
is seen as a strong DCN(3-2) source in the lower right of the region
figure~\ref{fig:c9e1e2}) (Liu et al., in prep.). However, there is
some localized, potentially elongated SiO emission that is spatially
coincident with C9A, which likely indicates that this core is already
forming a protostar \citep[similar to that present in C1-S][]{2016ApJ...821L...3T,2016ApJ...828..100F}. 
Most other cores do not show strong SiO(5-4)
emission. We discuss more details about the six strongest \ntdpns(3-2)
cores below.



In order to study basic core properties and, for the best cases of
C9A, B1A, H2A, B1B, B2A and F4A, compare to simple dynamical models,
we fit 2D elliptical Gaussians to the \ntdpns(3-2) integrated
intensity maps. We define an ellipse at the $3\sigma$ value of the 2D
Gaussian (Case 1), which is shown as the white, solid ellipses in the
\ntdpns(3-2) columns of Figs. \ref{fig:rowfigstart} to
\ref{fig:rowfigend}. These fitting results are listed in Table
\ref{tab:dynam} for the best six cores. Note that the 2D Gaussian
center is sometimes slightly different from the intensity weighted
core center defined earlier (Table \ref{tab:results}). Note also that
the \ntdpns(3-2) 0th-moment image, e.g., for C9A, has an rms of 13 mJy
per 1.5\arcsec$\times$1.0\arcsec beam \kmsns, roughly corresponding to
23 mJy per 2.3\arcsec$\times$2.0\arcsec \kmsns, which is about twice
the rms of C1 region rms reported by T13.  Thus the core sizes
reported here are likely to be systematically somewhat smaller than if
they had been measured with the same sensitivity used by T13. Still,
we will use these ellipses for our following study of core
dynamics. For C9A, B1A, H2A, B1B, B2A and F4A we also consider their
properties on the scale of larger ellipses (Case 2), shown by the
dashed ellipses in Figs.~\ref{fig:rowfigstart} to
\ref{fig:rowfigend}). These geometries are based on total \ntdpns(3-2)
intensities: they cover most of the \ntdp flux within the 2$\sigma$
contours from the full integration images. These ellipses also cover
most of the 1.3~mm continuum flux that is seen to be associated with
these cores. These core properties are shown inside square brackets in
Table~\ref{tab:dynam}.
We note that given these uncertainties in defining core sizes, we
  have not attempted to derive deconvolved sizes. As discussed above,
  our Case 1 radii are probably lower limits, and deconvolution would
  have only a very minor effect on the derived Case 2 radii.

In the following sections, we study the core dynamics, following the
methods of T13.  We focus on 6 cores C9A, B1A, H2A, B1B, B2A, and
F4A. The first four of these are the top four in terms of \ntdpns(3-2)
flux of the new cores presented in this paper (see
Table~\ref{tab:results}). They all have clear associations with 1.3~mm
continuum structures, which will be the preferred method of estimated
masses. B2A and F4A have relatively weak \ntdpns(3-2) fluxes, but also
show quite good correspondence with continuum sources.

\subsection{Core Kinematics}\label{subsec:corekinematic}



\begin{figure*}[htb!]
\epsscale{1.}
\plotone{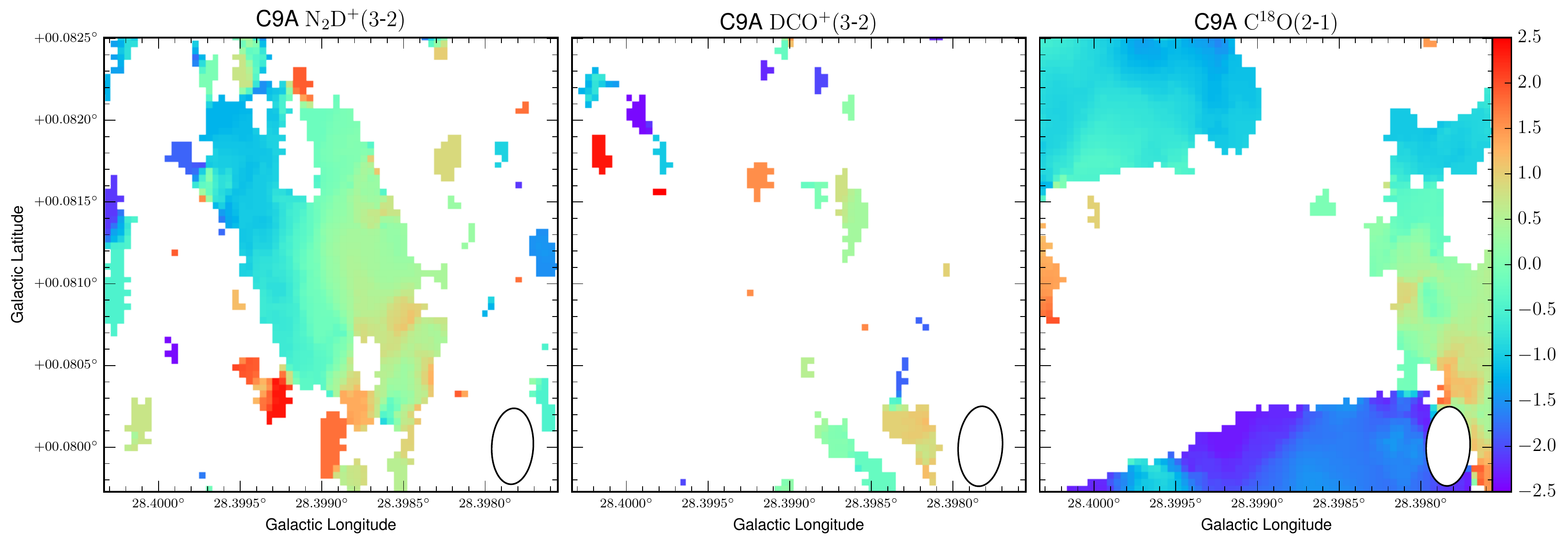}\\
\plotone{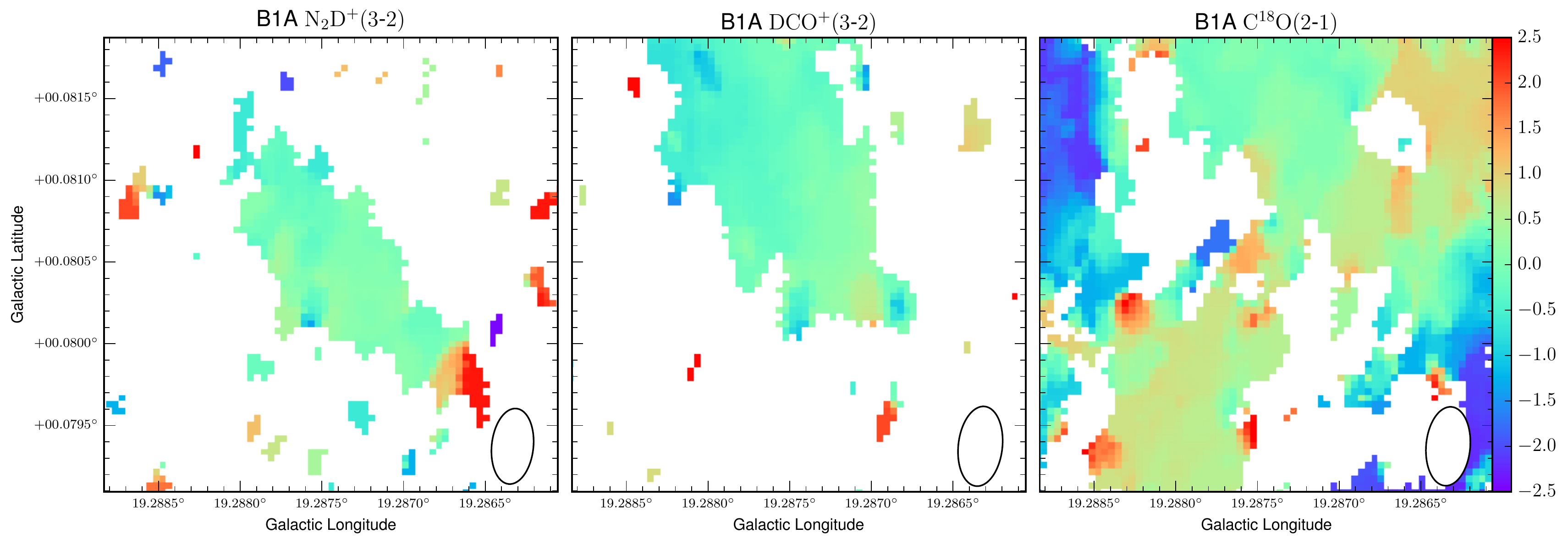}\\
\plotone{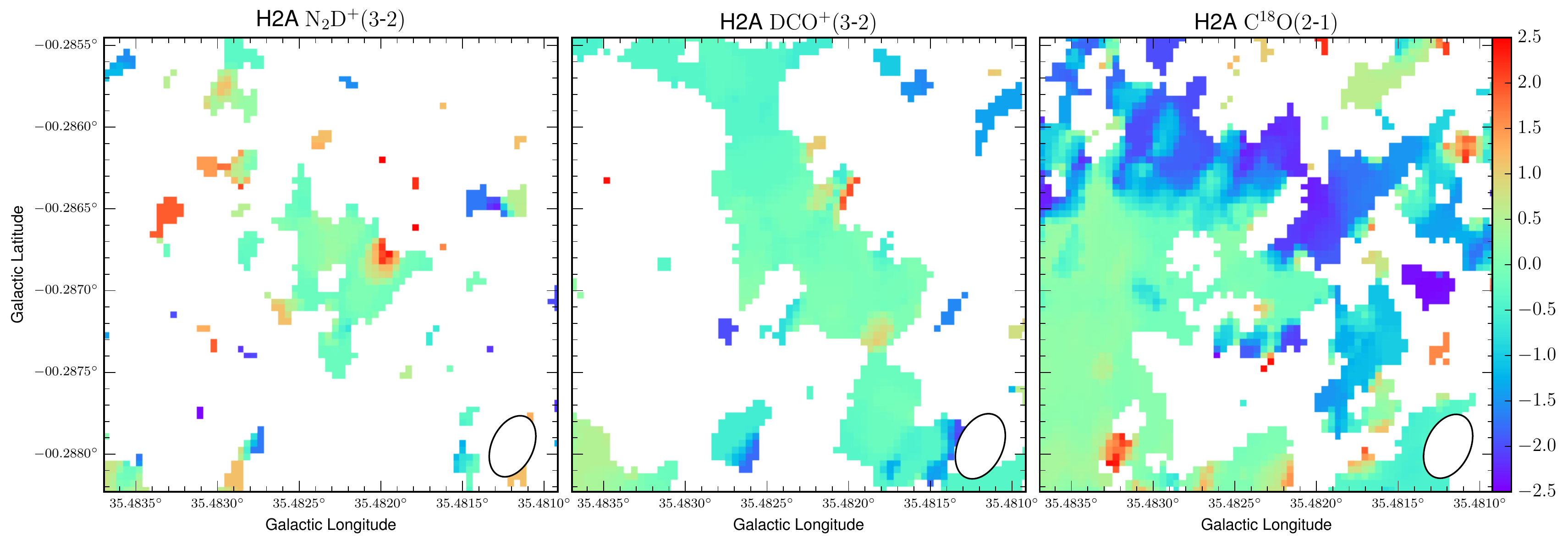}\\
\plotone{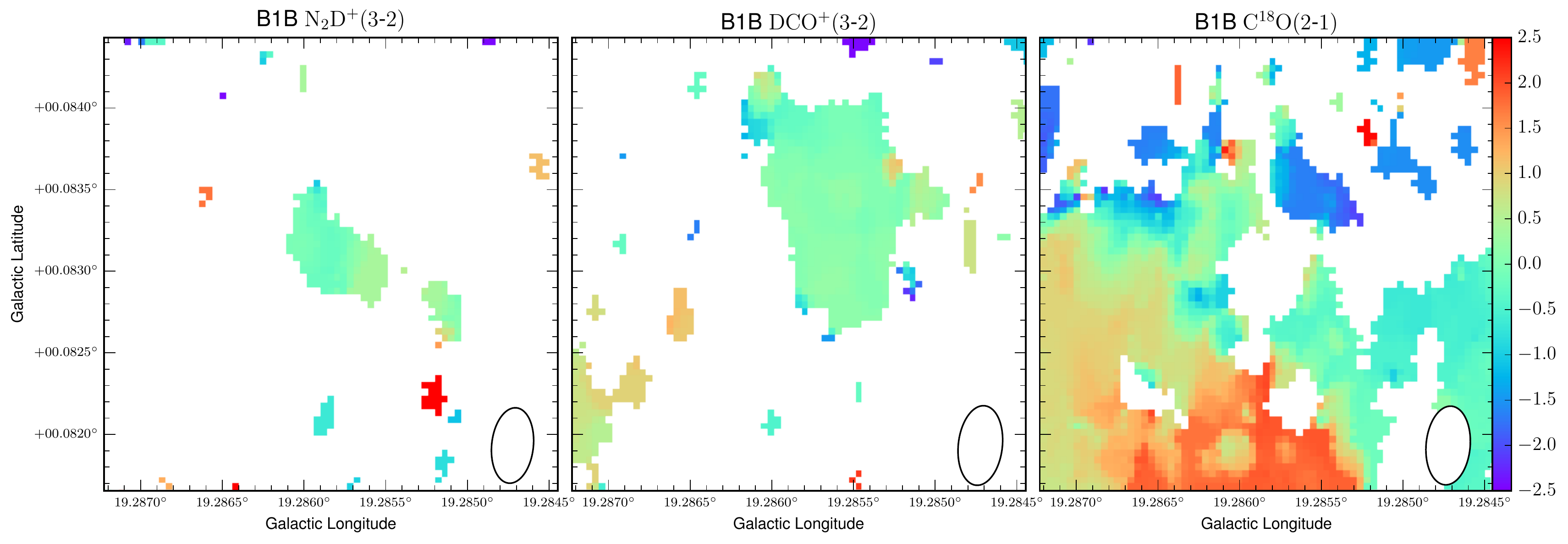}
\caption{
1st-moment (intensity weighted $v_{\rm LSR}$) maps of the top four
ranked \ntdpns(3-2) cores C9A (1st row), B1A (2nd row), H2A (3rd row), B1B (4th row),
in \ntdp (1st column), \dcop (2nd column), \ceio (3rd column).  The
calculation is within a 5 \kms range centered on \ntdpns(3-2) $v_{\rm
  LSR}$ (Table \ref{tab:results}).  Only voxels with $\geq 3 \sigma$
intensity are considered and shown in color.  The color scale is
linear from -2.5 \kms to +2.5 \kmsns.  The synthesized beam is shown
in the lower-right corner.
\label{fig:mom1}}
\end{figure*}

\begin{figure*}[htb!]
\epsscale{1.}
\plotone{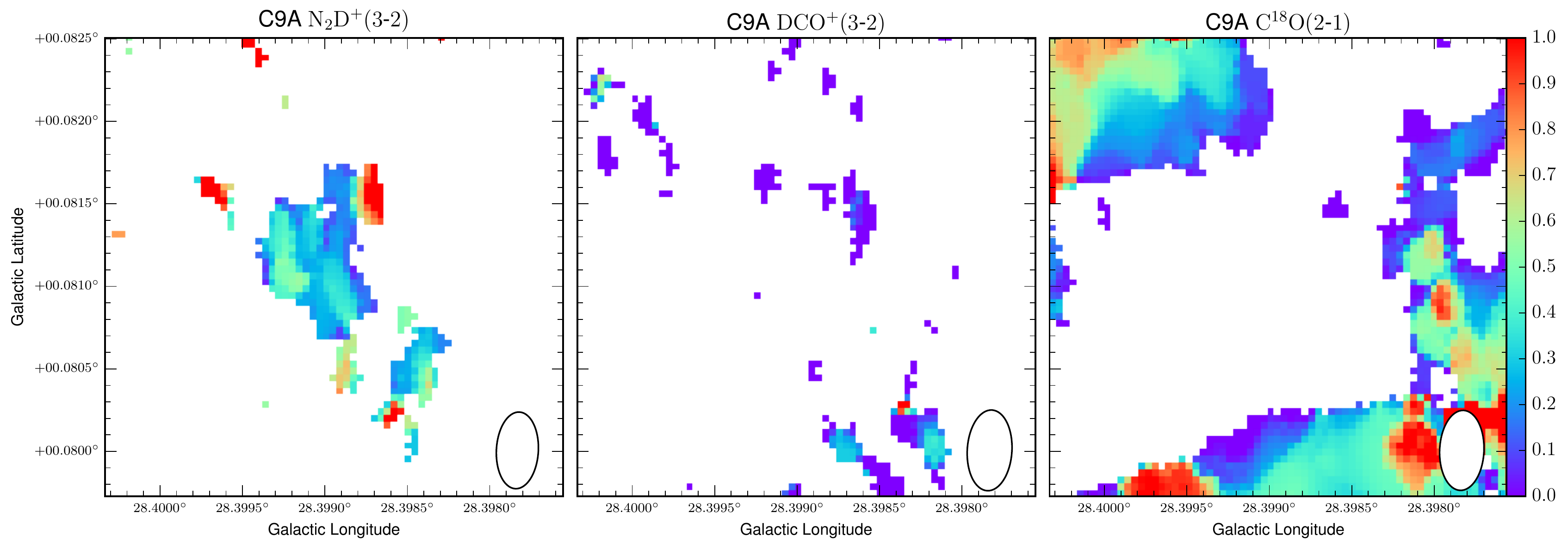}\\
\plotone{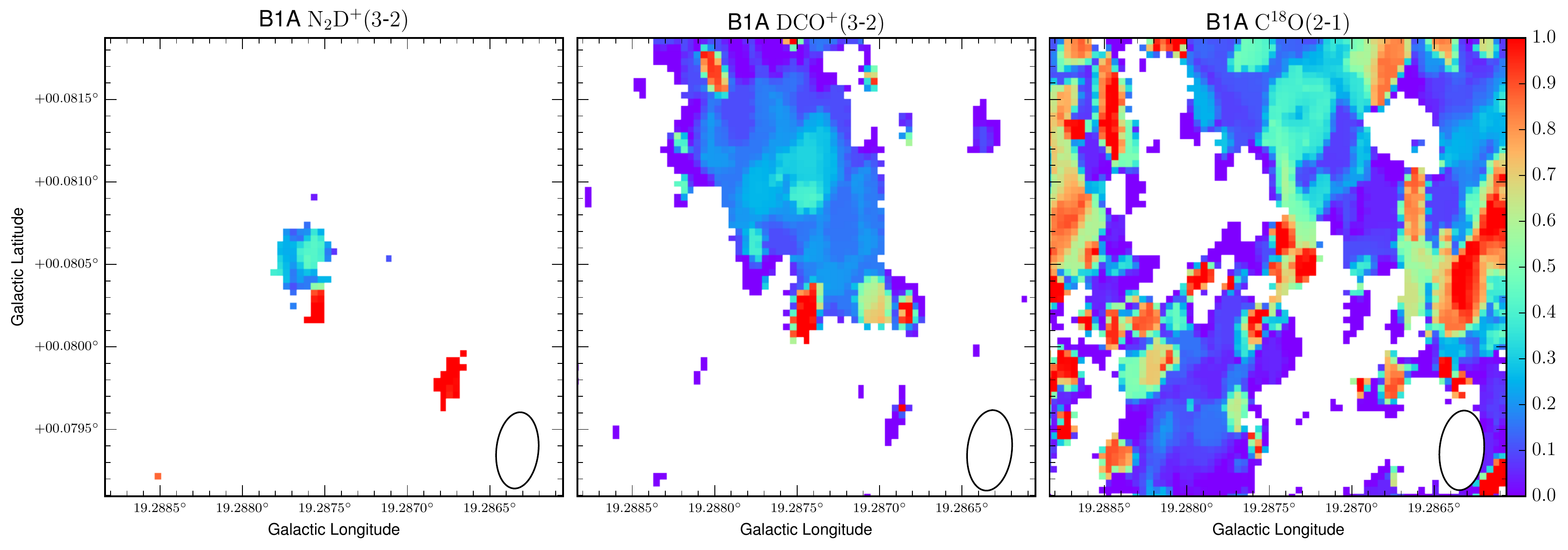}\\
\plotone{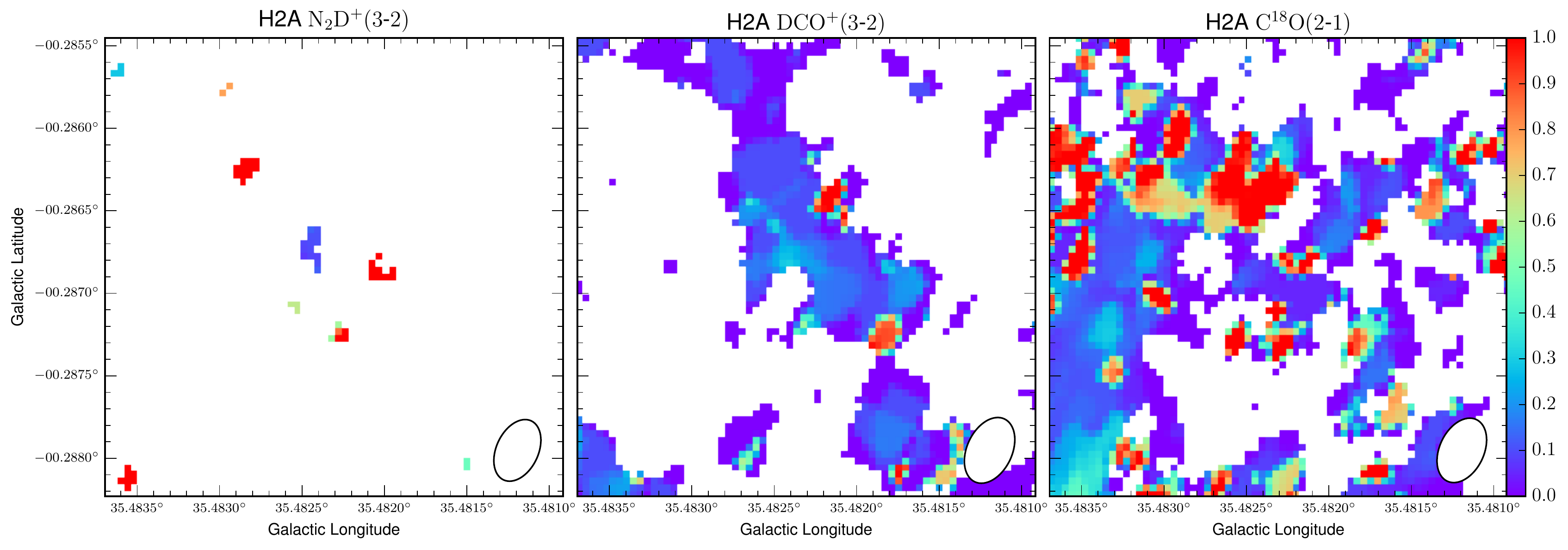}\\ 
\plotone{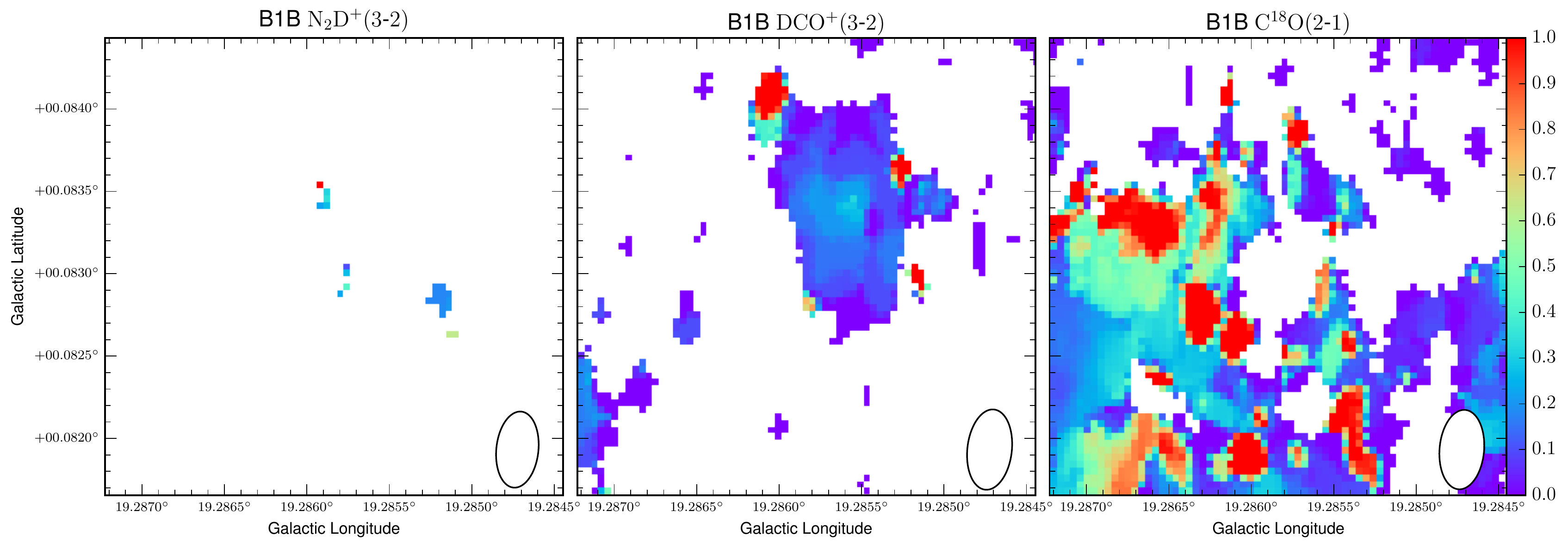}  
\caption{
2nd-moment (intensity weighted velocity dispersion) maps of the top
four ranked \ntdpns(3-2) cores C9A (1st row), B1A (2nd row), H2A (3rd
row), B1B (4th row), in \ntdp (1st column), \dcop (2nd column), \ceio (3rd column).
The calculation is within a 5 \kms range centered on \ntdp $v_{\rm
  LSR}$ (Table \ref{tab:results}).  Only voxels with $\geq$ 3$\sigma$
intensity are considered and shown in color.  For the \ntdp data, a
0.242 \kms contribution to the dispersion from the main group of
hyperfine structures is subtracted off in quadrature. However, we see
in some regions the 2nd-moment pixels shrink from the 1st-moment
images. This is mainly due to the fact that the velocity span is
narrower than 0.242 \kmsns. In these areas we are limited by the
sensitivity.  The color scale is linear from 0 to 1.0 \kmsns.  The
synthesized beam is shown in the lower-right corner.
\label{fig:mom2}}
\end{figure*}

Only the top four ranked \ntdpns(3-2) cores are significantly larger
than the beam. For these, Figure \ref{fig:mom1} shows the 1st-moment
maps of this species, along with \dcopns(3-2) and \ceions(3-2).
Only voxels with $\geq3 \sigma$ detection are considered. 
C9A shows a strong velocity gradient in \ntdpns(3-2): in the upper
left part of the image the mean velocities are $\simeq -1.5$~\kmsns,
while in lower right they are $\simeq +0.5\:$\kmsns. The core diameter
is about 0.1~pc, so the velocity gradient is
$\sim$20~\kms~pc$^{-1}$. Given the bimodal morphology of the
0th-moment map, it is possible that we are seeing two \ntdpns(3-2)
cores in the process of merging. The other cores, B1A and H2A, do not
show such large velocity gradients.



Figure \ref{fig:mom2} shows the 2nd-moment (velocity dispersion) maps
of C9A, B1A and H2A in \ntdpns(3-2), \dcopns(3-2) and \ceions(3-2).
Again, only voxels with $\geq 3 \sigma$ detection are considered. A
0.242~\kms effective dispersion from the \ntdpns(3-2) main hyperfine
group is subtracted in quadrature from its maps (regions that then
have a negative result are shown as blank: here the sensitivity is
probably too low to obtain a good measure of the velocity dispersion;
note also this effect leads to the appearance of artificial low
dispersion halos around the main features).
The C9A \ntdpns(3-2) 2nd-moment map shows a higher velocity dispersion
on its left-hand side compared to its right. Again, this may argue in
favor of the two merging cores scenario. 

\begin{figure*}[htb!]
\epsscale{.7}
\plotone{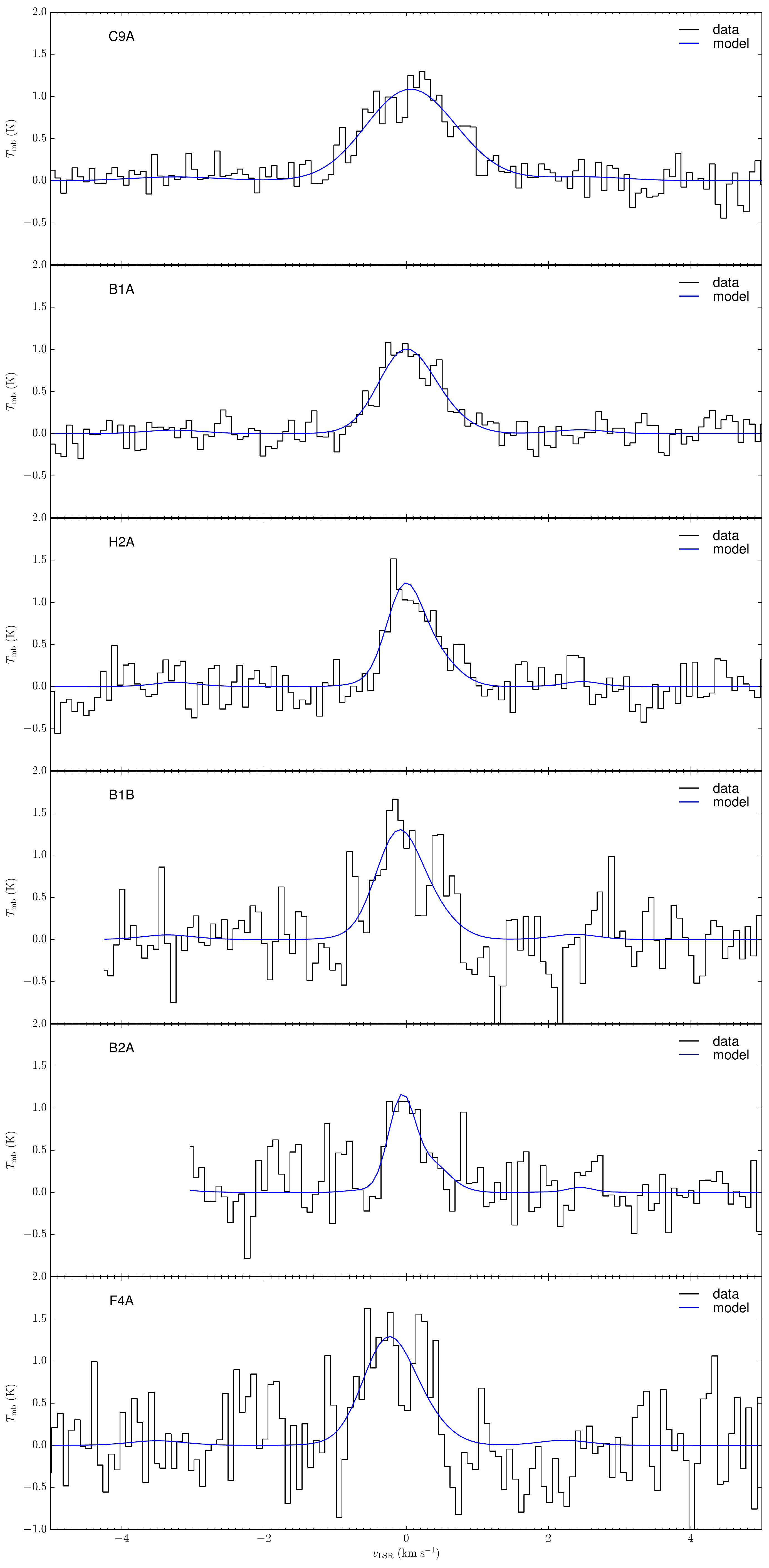}
\caption{
\ntdpns(3-2) core spectra (black) and their HFS fit results (blue).  All
spectra are in the rest frame of their centroid velocity $V_{\rm
  LSR,N_2D^+}$, derived from the HFS fit. The core names are labeled
on the top left.  Note that the spectra shown are binned to have 0.08
\kms spectral resolution, while the HFS fit results are drawn from
their maximum velocity resolution of 0.04 \kmsns.
\label{fig:specstart}}
\end{figure*}

Within the defined Case 1 ellipses of C9A, B1A, H2A, B1B, B2A, and
F4A, we extract total flux density from each channel (full velocity
resolution of 0.04~\kmsns) and convert it to main beam temperature,
\tmbns, to derive the total spectrum as a function of $v_{\rm
  LSR}$. These are shown in Figure~\ref{fig:specstart}. We then fit
the \ntdpns(3-2) line with full blended hyperfine components,
utilizing the HFS method of the CLASS software package
\footnote{http://www.iram.fr/IRAMFR/GILDAS}, assuming the line is
optically thin (this was shown to be good approximation in the cores
studied by T13). The resulting centroid velocity $V_{\rm LSR,N_2D^+}$
and 
1D velocity dispersion $\sigma_{\rm N_2D^+,obs}$
are listed in Table \ref{tab:dynam}.
The velocity dispersions derived by this method range from about 0.2
to 0.6~\kmsns.
However, to estimate the total velocity dispersion that is needed for
the dynamical analysis, we subtract the \ntdp thermal component from
its velocity dispersion in quadrature (assuming 10~K temperature), and
add back the core sound speed in quadrature, assuming a mean particle
mass $\mu = 2.33 m_p$ and gas temperature of 10$\pm$3~K, following
T13. These results, listed as $\sigma_{\rm N_2D^+}$ in
Table~\ref{tab:dynam}, are slightly larger, ranging from 0.26 to
0.61~\kmsns.


\subsection{Core Masses}\label{subsec:coremass}

Following the methods of T13, we measure core masses, $M_c$, within
the Case 1 and Case 2 ellipses in two ways. First we use the MIREX map
of BT12. As a first MIREX-based estimate we use the total mass surface
density in the map integrated over the area of the core to yield a
mass, $M_{c,{\rm max}}$. As a second MIREX-based estimate we account
for the clump mass surface density, $\Sigma_{\rm cl}$, in an
elliptical annulus around the core (from $R_c$ to $2R_c$ in the case
of a circular core), and then subtract off this contribution to
$\Sigma$ within the core ellipse, to then yield a mass, $M_{c,{\rm min}}$.

As discussed by T13, depending on the 3D structure of the core and
clump, there can be large differences between $M_{c,{\rm max}}$ and
$M_{c,{\rm min}}$. In some favorable situations where $\Sigma_{\rm
  cl}$ is relatively small, then both methods will yield similar mass
estimates. However, there can be cases where the surrounding ``clump''
has a higher inferred mass surface density and so $M_{c,{\rm min}}$ is
formally negative, i.e., this mass estimate is not well defined. In
fact this situation arises for C9A, since the \ntdpns(3-2) core is not
co-located with a MIREX $\Sigma$ peak. Another problem with the MIREX
based mass estimates is that the maps can suffer ``saturation'' in
high $\Sigma$ regions, i.e., where $\Sigma\gtrsim0.5\:{\rm
  g\:cm^{-2}}$ (see BT12). This leads to the MIREX based mass estimate
being an underestimate of the true core mass.

As a second method we calculate core masses from their 1.3~mm
continuum emission, $M_{c,{\rm mm}}$. We use equation (7) from T13 and
for consistency also assume a dust temperature of $T_d = 10 \pm
  3$~K (such cold temperatures are expected for highly deuterated
  cores; see also discussion of T13) and
adopt $\kappa_\nu = 5.95\times10^{-3}\:{\rm cm^2\:g}^{-1}$ \citep{1994A&A...291..943O}.
We adopt a 30\% uncertainty for $\kappa_\nu$. We do not attempt
to carry out clump envelope subtraction with this method, since the
ALMA observations tend to filter out the larger scale emission from
the clump. Since the uncertainties in temperature cause quite
asymmetric uncertainties in $\Sigma_{c,{\rm mm}}$ and $M_{c,{\rm
    mm}}$, we calculate lower and upper boundaries and put them in
sub- and super- scripts in Table~\ref{tab:dynam}. Overall, including
temperature, opacity and distance uncertainties, the total
uncertainty is about a factor of two in the mass estimation (see
detailed discussion in T13), however, because of the difficulties of
clump envelope subtraction in the MIREX mass estimates, we prefer the
mm continuum based mass estimate as our fiducial method.
We note that for a ``core'' contained within the synthesized beam
  size, at a distance of 5 kpc and with a dust temperature of 10 K,
  the 1$\sigma$ mass sensitivity is 0.51$\:M_\odot$.

With these methods we find that C9A is the most massive of the six
cores suitable for dynamical analysis, with $M_{c,{\rm mm}}\simeq
70\:M_\odot$ in the Case 1 (i.e., inner) ellipse. The other cores are
at least ten times smaller in mass. Considering the Case 2 (i.e.,
larger) ellipses, C9A rises in mass to $M_{c,{\rm mm}}\simeq
170\:M_\odot$. On these scales the other cores have about
$10\:M_\odot$.




\subsection{Core Dynamics}\label{subsec:coredynamic}

Following T13 and MT03, we consider virialized singular polytropic
quasi-spherical cores that have surfaces in approximate pressure
equilbrium with their surrounding clump environments. Such cores have
a radius $R_{\rm c,vir}$ and internal velocity dispersion $\sigma_{\rm
  c,vir}$ (assuming some contribution from large scale $B$-fields such
that the Alfv\'en Mach number in the core is unity).  The two
quantities are calculated following T13 equations (2) and (4), given
core mass $M_c$ and clump mass surface density $\Sigma_{\rm cl}$.  For
each core, we calculate the properties in 3 cases: (1) using the core
mass derived from MIREX map without envelope subtraction; (2) using
the core mass from MIREX map with envelope subtraction; (3) using the
core mass derived from the 1.3 mm continuum map. The clump mass
surface density is estimated from the MIREX map in the region from
$R_c$ to $2R_c$ (in the case of a circular core). Table
\ref{tab:dynam} lists the results for the six considered cores.

For our preferred mass estimates via 1.3~mm continuum emission we find
a mean ratio of the observed to predicted virial velocity dispersion
for the six cores of $\langle \sigma_{\rm N_2D^+}/\sigma_{\rm
  c,vir,mm}\rangle = 0.80$. This is very similar to the result found
by T13 of a ratio of 0.83 for the six cores they analyzed. It suggests
that this core population (of 12 cores) has properties that are
consistent with the virial equilbrium assumption of the MT03 Turbulent
Core model.

The most massive core, C9A, has a ratio of 0.71, which is modestly
subvirial. As discussed by T13, apparently subvirial cores may
indicate that stronger large scale magnetic fields are present: the
fiducial MT03 model assumes large scale $B$-fields are present that
would produce an Alfv\'en Mach number of unity for turbulence in the
core. T13 argued that stronger $B$-fields, $\sim 1$~mG, were needed in
the massive cores of their sample, especially C1-S, if the cores were
to be in virial equilibrium. We carry out a similar analysis here to
work out what $B$-field strengths are needed for virial equilbrium,
summarizing the results in Table~\ref{tab:3}. The magnetic field
strengths that are implied by the fiducial MT03 model are again
$\sim1$~mG. They need to be raised modestly to achieve precise virial
equilbrium. Also listed is the $B$-field strength, $B_{\rm c,crit}$,
needed to make the core mass equal to the magnetic critical mass
\citep{1992ApJ...395..140B}: these tend to be slightly higher again.

\begin{table*}
\centering
\scriptsize
\caption{Physical properties of the six ``best'' \ntdp cores.}\label{tab:dynam}
\begin{tabular}{cccccccc}
\hline {{Core property (\% error)}} & {{C9A}} & {{B1A}} & {{H2A}} & {{B1B}} & {{B2A}} & {{F4A}} & {{Average}}\\
\hline
$l$ ($^\circ$) & 28.39896 & 19.28747 & 35.48231 & 19.28584 & 19.30986 & 34.45843 & ... \\ 
 & [28.39896] & [19.28747] & [35.48231] & [19.28567] & [19.30983] & [34.45846] & ... \\ 
$b$ ($^\circ$) & 0.08113 & 0.08050 & -0.28681 & 0.08305 & 0.06689 & 0.25643 & ... \\ 
 & [0.08113] & [0.08050] & [-0.28681] & [0.08308] & [0.06699] & [0.25651] & ... \\ 
$\theta_c$ (\arcsec) & 1.94 & 1.65 & 1.42 & 0.707 & 0.745 & 0.606 & ... \\ 
 & [3.30] & [3.05] & [2.28] & [1.44] & [1.80] & [1.53] & ... \\ 
$e$ & 0.747 & 0.877 & 0.852 & 0.881 & 0.672 & 0.955 & ... \\ 
 & [0.904] & [0.866] & [0.781] & [0.00] & [0.00] & [0.866] & ... \\ 
P.A. ($^\circ$) & 16 & 42 & 45 & 58 & 83 & 85 & ... \\ 
 & [16] & [20] & [45] & [0] & [0] & [70] & ... \\ 
\hline
$d$ (kpc) (20\%) & 5.0 & 2.4 & 2.9 & 2.4 & 2.4 & 3.7 & ... \\ 
\hline
$R_{\rm c}$ (0.01~pc) (20\%) & 4.71 & 1.92 & 2.00 & 0.822 & 0.867 & 1.09 & ... \\ 
 & [7.99] & [3.55] & [3.20] & [1.67] & [2.09] & [2.74] & ... \\ 
\hline
\hline
$V_{\rm LSR,N_2D^+}$ ${\rm (km\:s^{-1})}$ & 78.40$\pm$0.02 & 27.60$\pm$0.02 & 45.50$\pm$0.02 & 26.60$\pm$0.04 & 25.50$\pm$0.03 & 57.10$\pm$0.05 & ... \\ 
 & [78.40$\pm$0.03] & [27.50$\pm$0.02] & [45.50$\pm$0.03] & [26.70$\pm$0.04] & [25.50$\pm$0.04] & [57.10$\pm$0.11] & ... \\ 
$\sigma_{\rm N_2D^+,obs}$ ${\rm (km\:s^{-1})}$ & 0.579$\pm$0.028 & 0.351$\pm$0.025 & 0.245$\pm$0.029 & 0.296$\pm$0.068 & 0.180$\pm$0.029 & 0.329$\pm$0.053 & ... \\ 
 & [0.579$\pm$0.028] & [0.351$\pm$0.025] & [0.245$\pm$0.029] & [0.296$\pm$0.068] & [0.180$\pm$0.029] & [0.329$\pm$0.053] & ... \\ 
$\sigma_{\rm N_2D^+,nt}$ ${\rm (km\:s^{-1})}$ & 0.576$\pm$0.028 & 0.347$\pm$0.025 & 0.239$\pm$0.030 & 0.291$\pm$0.069 & 0.172$\pm$0.030 & 0.325$\pm$0.053 & ... \\ 
 & [0.576$\pm$0.028] & [0.347$\pm$0.025] & [0.239$\pm$0.030] & [0.291$\pm$0.069] & [0.172$\pm$0.030] & [0.325$\pm$0.053] & ... \\ 
$\sigma_{\rm N_2D^+}$ ${\rm (km\:s^{-1})}$ & 0.606$\pm$0.028 & 0.395$\pm$0.026 & 0.304$\pm$0.029 & 0.347$\pm$0.062 & 0.255$\pm$0.029 & 0.375$\pm$0.049 & ... \\ 
 & [0.606$\pm$0.028] & [0.395$\pm$0.026] & [0.304$\pm$0.029] & [0.347$\pm$0.062] & [0.255$\pm$0.029] & [0.375$\pm$0.049] & ... \\ 
\hline
\hline
$\Sigma_{\rm cl}$ ${\rm (g\:cm^{-2})}$ (30\%) & 0.317 & 0.394 & 0.321 & 0.361 & 0.293 & 0.313 & ... \\ 
 & [0.312] & [0.321] & [0.292] & [0.362] & [0.273] & [0.288] & ... \\ 
\hline
$\Sigma_{\rm c,max}$ ${\rm (g\:cm^{-2})}$ (30\%) & 0.282 & 0.492 & 0.327 & 0.391 & 0.302 & 0.349 & ... \\ 
 & [0.308] & [0.438] & [0.326] & [0.437] & [0.297] & [0.334] & ... \\ 
\hline
$M_{\rm c,max}$ $(M_\odot)$ (50\%) & 9.34 & 2.70 & 1.95 & 0.394 & 0.338 & 0.616 & ... \\ 
 & [29.3] & [8.24] & [4.99] & [1.83] & [1.94] & [3.74] & ... \\ 
\hline
$n_{\rm H,c,max}$ $(10^5{\rm cm}^{-3})$ (36\%) & 6.17 & 26.5 & 16.9 & 49.1 & 35.9 & 33.1 & ... \\ 
 & [3.97] & [12.7] & [10.5] & [26.9] & [14.6] & [12.6] & ... \\ 
\hline
$\sigma_{\rm c,vir,max}$ ${\rm (km\:s^{-1})}$ & 0.514$\pm$0.064 & 0.398$\pm$0.050 & 0.348$\pm$0.044 & 0.241$\pm$0.030 & 0.220$\pm$0.027 & 0.259$\pm$0.032 & ... \\ 
 & [0.681$\pm$0.086] & [0.500$\pm$0.063] & [0.430$\pm$0.054] & [0.353$\pm$0.044] & [0.334$\pm$0.042] & [0.399$\pm$0.050] & ... \\ 
\hline
$\sigma_{\rm N_2D^+}/\sigma_{\rm c,vir,max}$ & 1.18$\pm$0.14 & 0.992$\pm$0.129 & 0.873$\pm$0.130 & 1.44$\pm$0.30 & 1.16$\pm$0.18 & 1.45$\pm$0.25 & 1.18$\pm$0.08 \\ 
 & [0.890$\pm$0.108] & [0.790$\pm$0.103] & [0.707$\pm$0.105] & [0.982$\pm$0.208] & [0.763$\pm$0.123] & [0.941$\pm$0.163] & [0.845$\pm$0.057] \\ 
\hline
$R_{\rm c,vir,max}$ (0.01~pc) & 4.02$\pm$1.03 & 1.94$\pm$0.49 & 1.83$\pm$0.46 & 0.774$\pm$0.198 & 0.796$\pm$0.204 & 1.04$\pm$0.26 & ... \\ 
 & [7.18$\pm$1.84] & [3.75$\pm$0.96] & [3.06$\pm$0.78] & [1.66$\pm$0.42] & [1.98$\pm$0.50] & [2.67$\pm$0.68] & ... \\ 
\hline
$R_c/R_{\rm c,vir,max}$ & 1.17$\pm$0.33 & 0.989$\pm$0.282 & 1.09$\pm$0.31 & 1.06$\pm$0.30 & 1.09$\pm$0.31 & 1.05$\pm$0.29 & 1.07$\pm$0.12 \\ 
 & [1.11$\pm$0.31] & [0.946$\pm$0.270] & [1.05$\pm$0.29] & [1.01$\pm$0.28] & [1.06$\pm$0.30] & [1.03$\pm$0.29] & [1.03$\pm$0.12] \\ 
\hline
\hline
$\Sigma_{\rm c,min}$ ${\rm (g\:cm^{-2})}$ (30\%) & - & 0.0979 & 6.37e-3 & 0.0303 & 8.68e-3 & 0.0368 & ... \\ 
 & - & [0.117] & [0.0342] & [0.0743] & [0.0243] & [0.0457] & ... \\ 
\hline
$M_{\rm c,min}$ $(M_\odot)$ (50\%) & - & 0.537 & 0.0380 & 0.0305 & 9.72e-3 & 0.0650 & ... \\ 
 & - & [2.20] & [0.522] & [0.311] & [0.159] & [0.511] & ... \\ 
\hline
$n_{\rm H,c,min}$ $(10^5{\rm cm}^{-3})$ (36\%) & - & 5.26 & 0.329 & 3.80 & 1.03 & 3.49 & ... \\ 
 & - & [3.39] & [1.10] & [4.58] & [1.20] & [1.72] & ... \\ 
\hline
$\sigma_{\rm c,vir,min}$ ${\rm (km\:s^{-1})}$ & - & 0.266$\pm$0.033 & 0.130$\pm$0.016 & 0.127$\pm$0.016 & 0.0905$\pm$0.0114 & 0.148$\pm$0.018 & ... \\ 
 & - & [0.359$\pm$0.045] & [0.245$\pm$0.030] & [0.227$\pm$0.028] & [0.179$\pm$0.022] & [0.243$\pm$0.030] & ... \\ 
\hline
$\sigma_{\rm N_2D^+}/\sigma_{\rm c,vir,min}$ & - & 1.49$\pm$0.19 & 2.34$\pm$0.34 & 2.74$\pm$0.57 & 2.82$\pm$0.45 & 2.54$\pm$0.43 & 2.38$\pm$0.18 \\ 
 & - & [1.10$\pm$0.14] & [1.24$\pm$0.18] & [1.53$\pm$0.32] & [1.43$\pm$0.23] & [1.55$\pm$0.26] & [1.37$\pm$0.10] \\ 
\hline
$R_{\rm c,vir,min}$ (0.01~pc) & - & 0.865$\pm$0.222 & 0.255$\pm$0.065 & 0.215$\pm$0.055 & 0.135$\pm$0.034 & 0.338$\pm$0.086 & ... \\ 
 & - & [1.94$\pm$0.49] & [0.991$\pm$0.254] & [0.687$\pm$0.176] & [0.565$\pm$0.145] & [0.987$\pm$0.253] & ... \\ 
\hline
$R_c/R_{\rm c,vir,min}$ & - & 2.22$\pm$0.63 & 7.84$\pm$2.24 & 3.82$\pm$1.09 & 6.42$\pm$1.83 & 3.22$\pm$0.92 & 4.70$\pm$0.65 \\ 
 & - & [1.83$\pm$0.52] & [3.23$\pm$0.92] & [2.44$\pm$0.69] & [3.71$\pm$1.06] & [2.78$\pm$0.79] & [2.80$\pm$0.36] \\ 
\hline
\hline
$S_{\rm 1.30mm}$ (mJy) & 31.3$\pm$0.8 & 9.63$\pm$0.69 & 5.52$\pm$0.61 & 10.3$\pm$0.3 & 4.36$\pm$0.31 & 1.96$\pm$0.26 & ... \\ 
 & [76.4$\pm$1.3] & [20.1$\pm$1.2] & [10.7$\pm$0.9] & [27.6$\pm$0.6] & [14.2$\pm$0.7] & [7.24$\pm$0.66] & ... \\ 
$S_{\rm 1.30mm}/\Omega$ (MJy/sr) & 112$\pm$2 & 47.6$\pm$3.4 & 37.1$\pm$4.1 & 274$\pm$7 & 107$\pm$7 & 71.9$\pm$9.6 & ... \\ 
 & [95.0$\pm$1.7] & [29.1$\pm$1.8] & [28.2$\pm$2.5] & [183$\pm$3] & [59.3$\pm$3.1] & [41.5$\pm$3.8] & ... \\ 
\hline
$\Sigma_{\rm c,mm}$ ${\rm (g\:cm^{-2})}$ & 2.10$_{1.24}^{4.21}$ & 0.895$_{0.530}^{1.79}$ & 0.698$_{0.413}^{1.40}$ & 5.15$_{3.05}^{10.3}$ & 2.01$_{1.19}^{4.03}$ & 1.35$_{0.800}^{2.71}$ & ... \\ 
 & [1.79$_{1.06}^{3.58}$] & [0.547$_{0.324}^{1.10}$] & [0.531$_{0.314}^{1.06}$] & [3.44$_{2.03}^{6.89}$] & [1.12$_{0.660}^{2.23}$] & [0.781$_{0.462}^{1.56}$] & ... \\ 
\hline
$M_{\rm c,mm}$ $(M_\odot)$ & 69.7$_{31.7}^{146}$ & 4.91$_{2.24}^{10.3}$ & 4.16$_{1.90}^{8.71}$ & 5.19$_{2.36}^{10.9}$ & 2.26$_{1.03}^{4.73}$ & 2.39$_{1.09}^{5.00}$ & ... \\ 
 & [170$_{77.6}^{357}$] & [10.3$_{4.69}^{21.6}$] & [8.11$_{3.70}^{17.0}$] & [14.4$_{6.55}^{30.1}$] & [7.29$_{3.32}^{15.3}$] & [8.73$_{3.98}^{18.3}$] & ... \\ 
\hline
$n_{\rm H,c,mm}$ $(10^5{\rm cm}^{-3})$ & 45.9$_{25.5}^{93.8}$ & 48.0$_{26.6}^{98.1}$ & 35.9$_{19.9}^{73.4}$ & 644$_{357}^{1320}$ & 239$_{133}^{489}$ & 128$_{71.0}^{261}$ & ... \\ 
 & [23.0$_{12.8}^{47.0}$] & [15.8$_{8.79}^{32.4}$] & [17.1$_{9.47}^{34.9}$] & [211$_{117}^{432}$] & [54.8$_{30.4}^{112}$] & [29.3$_{16.3}^{59.9}$] & ... \\ 
\hline
$\sigma_{\rm c,vir,mm}$ ${\rm (km\:s^{-1})}$ & 0.849$_{0.681}^{1.03}$ & 0.462$_{0.371}^{0.561}$ & 0.421$_{0.338}^{0.511}$ & 0.458$_{0.368}^{0.556}$ & 0.353$_{0.283}^{0.429}$ & 0.364$_{0.292}^{0.442}$ & ... \\ 
 & [1.06$_{0.849}^{1.28}$] & [0.528$_{0.424}^{0.641}$] & [0.486$_{0.390}^{0.590}$] & [0.592$_{0.475}^{0.718}$] & [0.465$_{0.373}^{0.565}$] & [0.493$_{0.396}^{0.599}$] & ... \\ 
\hline
$\sigma_{\rm N_2D^+}/\sigma_{\rm c,vir,mm}$ & 0.714$_{0.588}^{0.890}$ & 0.854$_{0.703}^{1.06}$ & 0.723$_{0.595}^{0.901}$ & 0.758$_{0.624}^{0.944}$ & 0.722$_{0.594}^{0.900}$ & 1.03$_{0.850}^{1.29}$ & 0.800$_{0.742}^{0.882}$ \\ 
 & [0.573$_{0.472}^{0.715}$] & [0.747$_{0.615}^{0.931}$] & [0.626$_{0.516}^{0.781}$] & [0.586$_{0.483}^{0.731}$] & [0.548$_{0.451}^{0.683}$] & [0.761$_{0.627}^{0.948}$] & [0.640$_{0.594}^{0.705}$] \\ 
\hline
$R_{\rm c,vir,mm}$ (0.01~pc) & 11.0$_{6.82}^{16.1}$ & 2.62$_{1.62}^{3.83}$ & 2.67$_{1.66}^{3.91}$ & 2.81$_{1.74}^{4.11}$ & 2.06$_{1.28}^{3.01}$ & 2.05$_{1.27}^{3.00}$ & ... \\ 
 & [17.3$_{10.7}^{25.3}$] & [4.20$_{2.61}^{6.14}$] & [3.90$_{2.42}^{5.72}$] & [4.67$_{2.90}^{6.83}$] & [3.83$_{2.38}^{5.61}$] & [4.08$_{2.53}^{5.97}$] & ... \\ 
\hline
$R_c/R_{\rm c,vir,mm}$ & 0.429$_{0.288}^{0.657}$ & 0.733$_{0.493}^{1.12}$ & 0.749$_{0.504}^{1.15}$ & 0.293$_{0.197}^{0.448}$ & 0.421$_{0.283}^{0.646}$ & 0.531$_{0.357}^{0.814}$ & 0.526$_{0.452}^{0.646}$ \\ 
 & [0.462$_{0.310}^{0.708}$] & [0.846$_{0.569}^{1.30}$] & [0.820$_{0.551}^{1.26}$] & [0.359$_{0.241}^{0.550}$] & [0.547$_{0.367}^{0.838}$] & [0.672$_{0.451}^{1.03}$] & [0.617$_{0.531}^{0.757}$] \\ 
\hline
\end{tabular}
\end{table*}

\begin{table*}
\centering
\begin{threeparttable}
\small
\caption{Dynamical properties of the six ``best'' \ntdp cores.}\label{tab:3}
\begin{tabular}{ccccccc}
\hline {{Core property (\% error)}} & {{C9A}} & {{B1A}} & {{H2A}} & {{B1B}} & {{B2A}} & {{F4A}}\\
\hline
$R_{\rm c}$ (0.01~pc) (20\%) & 4.71 & 1.92 & 2.00 & 0.822 & 0.867 & 1.09  \\ 
$\sigma_{\rm N_2D^+}$ ${\rm (km\:s^{-1})}$ & 0.606$\pm$0.028 & 0.395$\pm$0.026 & 0.304$\pm$0.029 & 0.347$\pm$0.062 & 0.255$\pm$0.029 & 0.375$\pm$0.049  \\ 
$\Sigma_{\rm cl}$ ${\rm (g\:cm^{-2})}$ (30\%) & 0.317 & 0.394 & 0.321 & 0.361 & 0.293 & 0.313  \\ 
$M_{\rm c,mm}$ $(M_\odot)$ & 69.7$_{31.7}^{146}$ & 4.91$_{2.24}^{10.3}$ & 4.16$_{1.90}^{8.71}$ & 5.19$_{2.36}^{10.9}$ & 2.26$_{1.03}^{4.73}$ & 2.39$_{1.09}^{5.00}$  \\ 
$\alpha_{\rm c}\equiv 5\sigma_{\rm N_2D^+}^2 R_{\rm c}/(G M_{\rm c,mm})^{a}$  & 0.287$_{0.124}^{0.636}$ & 0.702$_{0.299}^{1.56}$ & 0.514$_{0.211}^{1.14}$ & 0.220$_{0.0775}^{0.501}$ & 0.288$_{0.115}^{0.644}$ & 0.742$_{0.288}^{1.66}$  \\ 
$n_{\rm H,c,mm}$ $(10^5{\rm cm}^{-3})$ & 45.9$_{25.5}^{93.8}$ & 48.0$_{26.6}^{98.1}$ & 35.9$_{19.9}^{73.4}$ & 644$_{357}^{1320}$ & 239$_{133}^{489}$ & 128$_{71.0}^{261}$  \\ 
$t_{\rm c,ff}$ $(10^5{\rm yr})^{b}$ & 0.204$_{0.142}^{0.273}$ & 0.199$_{0.139}^{0.267}$ & 0.230$_{0.161}^{0.309}$ & 0.0544$_{0.0380}^{0.0730}$ & 0.0892$_{0.0624}^{0.120}$ & 0.122$_{0.0853}^{0.164}$  \\ 
$B_c$ $({\rm \mu G})$ ($m_A=1$) & 1220$_{903}^{1750}$ & 812$_{598}^{1160}$ & 541$_{393}^{780}$ & 2620$_{1800}^{3830}$ & 1170$_{842}^{1690}$ & 1260$_{899}^{1830}$  \\ 
$R_{\rm c,vir,mm}$ (0.01~pc) & 11.0$_{16.3}^{7.17}$ & 2.62$_{3.89}^{1.71}$ & 2.67$_{3.97}^{1.74}$ & 2.81$_{4.18}^{1.83}$ & 2.06$_{3.06}^{1.34}$ & 2.05$_{3.05}^{1.34}$  \\ 
$R_c/R_{\rm c,vir,mm}$  & 0.429$_{0.288}^{0.657}$ & 0.733$_{0.493}^{1.12}$ & 0.749$_{0.504}^{1.15}$ & 0.293$_{0.197}^{0.448}$ & 0.421$_{0.283}^{0.646}$ & 0.531$_{0.357}^{0.814}$  \\ 
$\sigma_{\rm c,vir,mm}$ ${\rm (km\:s^{-1})}$ & 0.849$_{0.681}^{1.03}$ & 0.462$_{0.371}^{0.561}$ & 0.421$_{0.338}^{0.511}$ & 0.458$_{0.368}^{0.556}$ & 0.353$_{0.283}^{0.429}$ & 0.364$_{0.292}^{0.442}$  \\ 
$\sigma_{\rm N_2D^+}/\sigma_{\rm c,vir,mm}$  & 0.714$_{0.588}^{0.890}$ & 0.854$_{0.703}^{1.06}$ & 0.723$_{0.595}^{0.901}$ & 0.758$_{0.624}^{0.944}$ & 0.722$_{0.594}^{0.900}$ & 1.03$_{0.850}^{1.29}$  \\ 
$\phi_{\rm B,vir}$ & 6.87$_{3.82}^{11.5}$ & 4.27$_{2.37}^{7.16}$ & 6.65$_{3.70}^{11.2}$ & 5.87$_{3.26}^{9.85}$ & 6.68$_{3.71}^{11.2}$ & 2.58$_{1.43}^{4.32}$  \\ 
$m_{\rm A,vir}$ & 0.519$_{0.383}^{0.772}$ & 0.711$_{0.506}^{1.18}$ & 0.529$_{0.390}^{0.791}$ & 0.573$_{0.419}^{0.874}$ & 0.528$_{0.389}^{0.788}$ & 1.08$_{0.704}^{3.36}$  \\ 
$B_{\rm c,vir}$ $({\rm \mu G})$ & 2350$_{1360}^{3670}$ & 1140$_{582}^{1830}$ & 1020$_{564}^{1620}$ & 4570$_{2500}^{7230}$ & 2220$_{1200}^{3530}$ & 1160$_{307}^{1980}$  \\ 
$B_{\rm c,crit}$ $({\rm \mu G})$ & 2830$_{1700}^{4730}$ & 1200$_{725}^{2010}$ & 939$_{565}^{1570}$ & 6930$_{4170}^{11600}$ & 2710$_{1630}^{4530}$ & 1820$_{1090}^{3040}$  \\ 
\hline
\end{tabular}
\begin{tablenotes}
\small
\item $^a$ Virial parameter \citep{1992ApJ...395..140B}.
\item $^b$ Core free-fall time, $t_{\rm c,ff} = [3\pi/(32G\rho_c)]^{1/2} = 1.38 \times 10^5 (n_{\rm H,c,mm}/10^5\:{\rm cm^{-3}})^{-1/2}\:{\rm yr}$.
\end{tablenotes}
\end{threeparttable}
\end{table*}

\subsection{Notes on Individual Cores}\label{subsec:individual}



\subsubsection{C9A}


C9A shows the strongest and most extended \ntdp structure
(Fig. \ref{fig:c9e1e2}). As discussed above, it shows complex
  kinematics, potentially indicative of two merging structures, which
  have now become connected in their \ntdp emission. C9A appears to
be in a very active, chaotic region (see Fig.~5 and 11). We see strong
detections of 1.3~mm continuum, especially a large filamentary
structure that contains C9A. The mm emission also shows evidence
  for potential fragmentation or localized heating within the C9A
  structure. We also see strong detections of DCO$^+$, DCN, \ceions
and CH$_3$OH from other sources in the vicinity, but not overlapping
with C9A. This core is close to
an HII region IRAS18402-0403, in the direction towards lower right in
the figure (mostly out of the FOV), following the mm continuum filament.
\citet{2010ApJ...721..222B} have studied this area (GLM4 clump) with
continuum and molecular line data (not including
\ntdpns). \citet{2009ApJ...696..268Z} studied this area (their P2
clump) and found two 1.3 mm continuum cores and an NH$_3$ core. They
correspond to the continuum sources to the south of C9A (about
7\arcsec away, see Figs. \ref{fig:c9e1e2} and \ref{fig:rowfigstart}).
We checked the Herschel 70 $\rm \mu m$ image in this area. There are
two sources. The larger source corresponds to the HII region. Next to
it is a smaller source about half the size, which corresponds to
the continuum source to the lower-right (Fig. \ref{fig:rowfigstart}). 
C9A is at the boundary of the smaller source.  

As discussed above, there is extended, strong SiO(5-4) emission in the
surroundings of C9A, likely driven from the strong continuum sources
to the lower left. However, C9A also has a weak, small SiO(5-4)
counterpart. This only shows up on the red side of the spectrum.  The
small SiO counterpart is detected from $\sim$ 82-89 \kmsns, while C9A
is detected from 77.05-80.35 \kmsns. No SiO emission on the blue side
is detected (the lower limit of velocity is $\sim$ 22.5 \kmsns). We
also check that there is no overlap between the \ntdp core and the SiO
connected structures in the PPV space, again using the Graph method.
It is unclear whether the small SiO(5-4) counterpart in the C9A \ntdp
core is part of the outflow from the external strong continuum source
or whether it indicates the presence of an already formed protostar
within the core \citep[similar to that present in C1-S][]{2016ApJ...821L...3T,2016ApJ...828..100F}.
This question will be investigated further by Liu
et al. (in prep.).

\subsubsection{B1A}

B1A appears to be in a more quiescent, isolated environment.  It shows
an elongated morphology, and has relatively weak 1.3~mm continuum
counterpart (compared to C9A).  
We have detected SiO(5-4) emission to the north of B1A outside the
continuum boundary. This does not appear to be originating from a
protostellar source within the B1A core. However, due to the limited
spectral coverage in the blue wing side, we cannot see if there is a
symmetric SiO structure on the other side of the B1A core.

\citet{2011ApJS..195...14S} have studied this region as one of the
BGPS sources \citep[][with 33" spatial
  resolution]{2011ApJS..192....4A}.  They observed HCO$^+$(3-2) and
\nthpns(3-2) lines with 30\arcsec\ resolution.  Their study is on the
clump scale. At this scale, they found a velocity dispersion of about
1.5 \kmsns, which is about a factor of 5 larger than our measurement
at core scale with \ntdpns(3-2).  We checked the Herschel 70~$\rm \mu
m$ image in this area. B1A has no 70 $\rm \mu m$ counterpart, and is
in a 70 $\rm \mu m$ dark area.

\subsubsection{H2A}

H2A appears to be in a quiescent environment.  It has no detectable
SiO(5-4) emission. It has a relatively weak 1.3~mm continuum
counterpart (compared to C9A). H2A's continuum shows an ``X-shaped''
structure.  We checked the Herschel 70 $\rm \mu m$ image in this area
finding that H2A has no corresponding 70 $\micron m$ source.

\subsubsection{B1B}

B1B is about 10\arcsec away from B1A (Fig. \ref{fig:b1b2c2}). It is
displaced in velocity from B1A by about 1~\kmsns. It has a strong
1.3~mm continuum counterpart (Fig. \ref{fig:rowfigstart}). There is a
MIR-bright source (visible as a ``hole'' in the MIREX map) nearby.

\subsubsection{B2A}

B2A corresponds to a continuum core that is at the end of a group of
1.3~mm continuum sources that also containts a MIR-bright source.

\subsubsection{F4A}

F4A corresponds to one of a pair of spatially adjacent 1.3~mm
continuum sources. The other source does not show significant \ntdpns
emission. Overall, this is a relatively quiescent environment.

\section{Discussion and Conclusions}\label{sec:dc}


We have carried out a survey of 32 IRDC clumps designed to detect
cores with strong \ntdpns(3-2) emission. Such cores may be massive
analogs of low-mass pre-stellar or early-stage protostellar
cores. This work follows on from the pilot study of T13, which
identified six such cores in 4 IRDC clumps. Our current survey has a
lower line sensitivity level than T13, but a similar 1.3~mm continuum
sensitivity. The spectral set-up includes several ancillary line
tracers, including DCO$^+$(3-2), DCN(3-2), C$^{18}$O(2-1) and
SiO(5-4). We have also utilized the MIREX maps of these regions
developed by \citet{2014ApJ...782L..30B} and BT12.

In order to process the larger number of target regions, we have
presented a new way to automatically identify \ntdpns(3-2) cores as
connected structures in PPV space using Graph theory methods.



In total 141 \ntdpns(3-2) core candidates were identified via these
automated methods, although many of the weakest sources are likely to
be noise fluctuations. The locations of these sources are identified
in our maps of the clump-scale regions. We have presented properties
of the strongest 50 cores, including their mean velocities and
velocity ranges, and their \ntdpns(3-2) line fluxes. We have presented
zoom-in maps of the top 15 of these cores and a dynamical analysis of
the best 6 amongst these sources.

The main results are the identification of the very massive (up to
$\sim 170\:M_\odot$) C9A \ntdpns(3-2) ``core,'' i.e., a connected
structure in ppv space, which shows complex structure and kinematics.
Several other $\sim10\:M_\odot$ cores are found. The \ntdpns(3-2)
velocity dispersions are consistent with the predictions of the
turbulent core model of MT03, based on quasi virial equilbrium of such
structures. Further follow-up work is needed to test for the starless
nature of these cores, especially examining outflow tracers.

The methods presented in this study should also be applied to larger
samples of clumps to identify their \ntdpns(3-2) core populations,
which may be key for understanding the origin of the stellar initial
mass function and the formation of star clusters.


\acknowledgments
We thank Brian Svoboda for helpful comments and suggestions.
JCT and SK acknowledge an NRAO/SOS grant and NSF grant AST1411527.
PC acknowledges the financial support of the European Research Council (ERC; project PALs 320620).
This paper makes use of the following ALMA data: 
ADS/JAO.ALMA\#2013.0.00806.S. ALMA is a partnership of 
ESO (representing its member states), NSF (USA) and NINS (Japan), 
together with NRC (Canada), NSC and ASIAA (Taiwan), 
and KASI (Republic of Korea), in cooperation with the Republic of Chile. 
The Joint ALMA Observatory is operated by ESO, AUI/NRAO and NAOJ. 
The National Radio Astronomy Observatory is a facility of the 
National Science Foundation operated under cooperative agreement by Associated Universities, Inc.

{\it Facilities:} \facility{ALMA}

\bibliography{ref}
\bibliographystyle{aasjournal}

\end{document}